\documentclass[a4paper]{amsart}
\usepackage{amsmath, amsthm, amssymb, amscd, mathrsfs, eucal,
  amsfonts, latexsym}
\usepackage{hyperref}

\newcommand{\p}{\pi}

\newcommand{\R}{{\mathbb R}}

\newcommand{\balder}{{\eth} } 
\newcommand{\scSHP}{\mbox{\rm \tiny{\tiny{SHP}}}}
\newcommand{\omM}{\omega_o}
\newcommand{\phM}{\phi_o{}}
\newcommand{\xM}{x_o}
\newcommand{\sM}{s_o}
\newcommand{\MM}{M_o}
\newcommand{\bB}{{\boldsymbol{\beta}}}
\theoremstyle{plain}
\swapnumbers
\newtheorem{Thm}{Theorem}[section]

\newtheorem{Lem}[Thm]{Lemma}

\newtheorem{Def}[Thm]{Definition}
\theoremstyle{remark}

\begin{document}

\title{}
\begin{center}
{\large \bf Local Thermal Equilibrium States and\\ Quantum Energy Inequalities}
${}$\\[16pt]
{\large \sc Jan Schlemmer${}^{1,2}$ and Rainer Verch${}^1$}
\\[16pt]
${}^1$ Institut f.\ Theoretische Physik, Universit\"at Leipzig,\\
Postfach 100 920, D-04009 Leipzig, Germany \\[6pt]
${}^2$ Max-Planck-Inst.\ f.\ Mathematics in the Sciences,\\
 Inselstr.\ 22, D-04103 Leipzig, Germany
\end{center}

\begin{abstract}
{In this paper we investigate the energy distribution of states of a linear 
scalar quantum field with arbitrary curvature coupling on a curved spacetime which fulfill some local thermality 
condition. We find that this condition implies a quantum  energy
inequality for these states, where the (lower)
energy bounds depend only on the local temperature distribution and are local and covariant (the
 dependence of the bounds other than on temperature is on
parameters defining the quantum field model, and on local quantities
constructed from the spacetime metric). Moreover, we
also establish  the averaged
null energy condition (ANEC) for such locally thermal states, under growth conditions
on their local temperature and under conditions on the free parameters
entering the definition of the renormalized stress-energy tensor. These results hold for
a range of curvature couplings including the cases of conformally coupled and minimally 
coupled scalar field.}
\end{abstract}
\maketitle
\section{Introduction}
One of the problems in quantum field theory is that the
space of states is enormous, and that it is difficult to
establish criteria which single out the states of interest
in various physical situations. While this problem is met
only in a mild form when considering quantum fields on
Minkowski spacetime, it becomes quite pressing when trying
to combine quantum field theory and gravity. A situation
where this is predominant is quantum field theory in curved
spacetime, where quantized matter fields propagate on a 
classical background spacetime which may be curved, and 
where the spacetime metric is not stationary -- spacetimes
of Friedman-Robertson-Walker type, for instance, are of particular interest
in this context as they are simple models for cosmological scenarios.
In trying to model the conditions of the early stages of the
universe, one would like to distinguish quantum field states which are, 
at least locally, not too far from thermal equilibrium, and for
which one can assign, at least locally, a temperature. 
It is not at all easy to arrive at a meaningful concept
of temperature for states of a quantum field on a generic
spacetime since the (global) notion of temperature makes reference to
a global inertial frame, and that is not available in the
presence of spacetime curvature. Another manifestation of
this circumstance is the observer-dependence of the concepts of
particle and temperature, as illustrated by the Fulling-Unruh,
and related effects \cite{Unruh,Fulling,GuidoLongo}. 

Nevertheless, following a proposal by Buchholz, Ojima and Roos
\cite{BuOjiRoo} and further investigated in \cite{Bu,BuSchl}, it is possible
to introduce a covariant concept of states which, at given
points $x$ in spacetime, look like thermal equilibrium states with respect to
a certain set, $S_x$, of reference-observables. Such states are called
{\em local thermal equilibrium (LTE) states} with respect to $S_x$.
While there is some leeway in the determination of $S_x$, it
is important that $S_x$ does not contain observables which
are sensitive to flux-like quantities,
but instead observables which correspond, in the situation of global thermal equilibrium
states defined with respect to an inertial frame (in the absence of
spacetime curvature), to intensive thermal quantities. For a (scalar)
quantum field $\phi(x)$, typical elements of $S_x$ are
the Wick-square $:\phi^2:(x)$ of the field and its so-called
``balanced derivatives''. We will discuss the concept of
LTE states further in the next section. However, it is important to mention
that for an LTE state $\omega$ of the massless linear scalar field on a generic 
(globally hyperbolic) spacetime, the expectation value of the Wick-square
of the field at any spacetime point $x$,
\begin{equation}
{\tt T}^2(x) = \langle :\phi^2:(x) \rangle_\omega \equiv \omega(:\phi^2:(x))
\end{equation}
equals, up to a constant, the square of the absolute temperature
of the state at $x$. (A similar statement holds for the linear scalar field with
positive mass parameter.) 

Quite clearly, the local thermodynamic properties of states are linked
with the local energetic properties of states, and investigation of that relation
is the topic of the present article. A quantity of prominent interest
in quantum field theory in curved spacetime -- and with relevance
to questions in cosmology
 -- is the expectation value of the stress-energy
tensor, $\langle T_{ab}(x)\rangle_\omega$, in a state $\omega$
for a quantum field on a curved spacetime. It becomes particularly important
when considering the semiclassical Einstein equations (in geometric units),
\begin{equation}
G_{ab}(x) = 8\pi ( T_{ab}^{class}(x) + \langle T_{ab}(x)\rangle_\omega)\,,
\end{equation}
where $G_{ab}(x)$ is the Einstein tensor of the spacetime geometry
at spacetime point $x$ and $T_{ab}^{class}$ is the stress-energy tensor of
macroscopically modelled  (classical) matter. If the gravitational curvature effects caused by 
the macroscopic energy/matter distribution are very high, this may 
induce quantum field theoretical ``particle creation effects'', as a result
of which $\langle T_{ab}(x)\rangle_\omega$ may turn out as a significant
correction to the macroscopic matter distribution, depending on the quantum
state $\omega$. One may expect that this is a realistic
scenario as nowadays the Casimir force puts limits to the design of
devices in micro- and nano-technology \cite{CasimEng1,CasimEng2}, and this effect can be seen in
a similar vein. 

One of the interesting features of the expectation value of stress energy is
that the energy density seen by an observer travelling on a timelike geodesic
$\gamma$ with tangent vector $v^a$ at $x$, 
$\langle T_{ab}(x)\rangle_\omega v^{a}v^{b}$,
is unbounded above {\em and} below as $\omega$ ranges over the set of all
states $\omega$ (for which the expectation value of stress-energy at any
spacetime point $x$ can be reasonably defined). This is a long known
feature of quantum field theory (see \cite{EpGlaJaf}) and is in contrast to the behaviour
of macroscopic matter which can -- with good motivation -- usually be assumed
to satisfy one of the classical energy conditions, like the weak energy condition,
which means $T_{ab}^{class}(x)v^{a}v^{b} \ge 0$, i.e.\ the energy density
seen by any observer is always positive at any spacetime point $x$.  

Energy positivity conditions like the (pointwise) weak energy conditions play
an important role in the derivation of singularity theorems 
\cite{HawkEllis,WaldGR}. One consequence
of energy positivity conditions when plugged into Einstein's equations is
that gravitational interaction is always attractive. Negative energies, in
contrast, would be affected by a repelling gravitational interaction. This
could, a priori, lead to solutions of Einstein's equations exhibiting very 
strange spacetime geometries, such as spacetimes with closed timelike curves,
wormholes or ``warpdrive scenarios'' \cite{MorThorYurt,Alcub}. 
Moreover, concentration of a vast amount
of negative energies and their persistence over a long duration could lead to
violations of the second law of thermodynamics.

Motivated by the latter point, L.\ Ford has proposed that physical states of 
quantum fields in generic spacetimes should not permit arbitrary concentration
of large amounts of negative energy over a long duration \cite{Ford}. Such limitations on
physical quantum field states have come to be called {\em quantum energy inequalities
(QEIs)}. Let us explain this concept in greater detail. Suppose that $\phi(x)$ is
a quantum field on a generic spacetime. (Actually, only smeared quantum field 
quantities like $\phi(f) = \int f(x) \phi(x)\, d{\rm vol}(x)$ define proper quantum 
field operators operators; 
$d{\rm vol}(x)$ denotes the metric induced spacetime volume form and $f$
a smooth, compactly supported test-function. The quantum field
can be of general spinor- or tensor type, but we suppress any corresponding 
indices here.)  Then let $\mathcal{L}$ be a set of states of the quantum
field such that the expectation value of the stress-energy tensor,
$\langle T_{ab}(x)\rangle_\omega$, is defined for each $\omega \in \mathcal{S}$
at each spacetime point $x$. We will furthermore suppose that this quantity is
continuous in $x$ for each $\omega \in \mathcal{L}$. Under these assumptions, we
say that the set of quantum field states $\mathcal{L}$ fulfills a 
QEI with respect to $\gamma$ if
\begin{equation}
 \int h^2(t)\langle T_{ab}(\gamma(t))\rangle_\omega \dot{\gamma}{}^a(t)\dot{\gamma}{}^b(t) \,dt
 \ge q(\gamma,h)
\end{equation}
holds for all smooth  (or at least $C^2$) real functions $h$ having compact support on the
(open) curve domain, with a constant $q(\gamma,h) > -\infty$; the
constant may depend on the curve $\gamma$ and the weighting function $h$,
but is required to be independent of the choice of state $\omega \in
\mathcal{L}$. 

There is a limiting case of a QEI: If $\gamma$ is a complete (lightlike
or null) geodesic, then one says that a set of states $\mathcal{L}$
fulfills the averaged null energy condition (ANEC) if
$$ \liminf_{\lambda \to 0+}\, \int h^2(\lambda t) 
\dot{\gamma}{}^a(t) \dot{\gamma}^b(t)\,\langle T_{ab}(\gamma(t))\rangle_\omega \,dt
\ge 0$$
holds for all states $\omega \in \mathcal{L}$. 
Conditions of such form (and related conditions, see \eqref{eins1}), if valid for all complete null
geodesics, allow conclusions about focussing of null geodesics for solutions
to the semiclassical Einstein equations similar to that resulting from a pointwise
null energy condition \cite{GFTipler,GFBorde,GFRoman, YurWal}. 
(See also the beginning of Section 4.)
Thus, the ANEC is a key property for deriving singularity theorems for 
solutions to the semiclassical Einstein equations.

Quantum energy inequalities have been investigated extensively for quantum fields 
subject to linear field equations in the recent years, and there is now a wealth
of results in this regard. We refer to  the reviews by Fewster and by Roman \cite{RevRoman,RevFewster}
for representative lists of references. Important to mention, however, is
the fact that for many linear fields, like the minimally coupled scalar field, the Dirac field
and the electromagnetic field, it could be shown that the set of Hadamard states fulfills
a QEI with respect to timelike curves $\gamma$ in generic globally hyperbolic spacetimes \cite{GenWlKG,GenWlDirac,GenWlEM}. Hadamard states are regarded as physical states in
quantum field theory in curved spacetime, and expectation values of the stress-energy
tensor at any given spacetime point are well-defined for these states (up to finite
renormalization ambiguities), cf. \cite{WaldQFT} for discussion. There is also 
an intimate relation between QEIs, the Hadamard condition and thermodynamic properties
of linear quantum fields \cite{CJF-RV}. It has been shown that QEIs put strong limitations
on the possibility of solutions to the semiclassical Einstein's equations to allow
exotic spacetime scenarios such as wormholes or warpdrive 
\cite{QEIvsWH1, QEIvsWH2, QEIvsWarp}. 
It is also worth mentioning two other recent results. First, it has been shown that
the non-minimally coupled linear scalar field on any spacetime violates QEIs for the
class of Hadamard states; nevertheless, the class of Hadamard states fulfills
in this case weaker bounds, called ``relative QEIs'', cf. \cite{FewOst} for results and
discussion. Secondly, one is interested in lower bounds $q_\gamma(h)$ which depend
(apart from renormalization constants entering the definition of expectation value
of the stress energy tensor) only on the underlying spacetime geometry in a local and
covariant manner, and one also aims at making this dependence as explicit as possible. Considerable progress on this issue, for the case of the minimally coupled linear 
scalar field on globally hyperbolic spacetimes, has been achieved in \cite{FewSmith}.

In the present article, we will derive QEI-like bounds on sets of LTE-states of
the non-minimally coupled linear scalar field $\phi(x)$ on generic globally hyperbolic spacetimes.
More precisely, we consider LTE states $\omega$ whose thermal function
$\vartheta^\omega(x) = \langle : \phi^2:(x) \rangle_\omega$ is bounded by some 
constant ${\tt T}_0^2$ (corresponding to a maximal squared temperature) and we will show that
there are upper and lower lower bounds for the averaged energy density
$$ \int_{-\infty}^\infty \eta(\tau) v^av^b \langle T_{ab}(\gamma(\tau)) \rangle_\omega \,d\tau \,,$$
averaged against a $C^2$-weighting function $\eta \ge 0$ with compact support along
any causal geodesic $\gamma$ with affine parameter $\tau$ and tangent 
$v^a = \dot{\gamma}{}^a$. The lower bound depends only on ${\tt T}^2_0$, the geodesic
$\gamma$ and $\eta$, while the upper bound depends additionally on local tetrads entering
into the definition of LTE states. The lower bound is therefore state-independent
within each set of LTE states $\omega$ with a fixed maximal value of $\vartheta^\omega$.
The bounds depend on the spacetime geometry in a local covariant manner which, together with
their dependence on ${\tt T}_0^2$, we will make explicit. This result holds for all values
of curvature coupling $\xi$ in the field equation \eqref{fieldeqn}, and upon averaging
along causal geodesic, not only those which are timelike. Hence, the result is not
immediate from know quantum energy inequalities for Hadamard states, as these are violated in
general for non-minimally coupled fields \cite{FewOst}, and upon averaging along null geodesics
\cite{FewRoNull}. Furthermore, we will show that the ANEC holds for LTE states $\omega$ of the
quantized linear scalar field with curvature couplings $0 \le \xi \le 1/4$, provided that the
growth of the thermal function $\vartheta^\omega$ along the null geodesics $\gamma$ fulfills
certain bounds. Despite the fact that we have to assume that the LTE states we consider
are Hadamard states -- in order to have a well-defined, local covariant expression of 
expected stress-energy for these states -- our derivation of QEIs and ANEC makes no further use
of the Hadamard property but uses only properties of LTE states. Therefore, one may expect that,
in principle, similar results could be derived for LTE states of interacting quantum fields.
This prospect can actually be seen as one of our motivations in view of the fact that
quantum energy inequalities seem to be very difficult to obtain (if valid at all)
 for very general sets of
states in interacting quantum field theory, and that, on the other hand, one may argue that
only special classes of states are of physical interest. We will come back to this point in
Sec.\ 6.

This article is organized as follows. We will discuss the concept of LTE states, as
far as needed for our purposes, in Section 2. In Section 3 we derive upper and lower
bounds for the geodesically averaged expectation values of energy density for LTE
states. The validity of  ANEC for certain LTE states will be
studied in Section 4. 
In Sec.\ 5 we indicate that the results of Secs.\ 3 and 4 hold also for a 
more general notion of LTE states.
 We conclude with
discussion and outlook in Sec.\ 6.

\section{Local Thermal Equilibrium States}

The system under investigation in the present article is the non-minimally
coupled linear scalar field on globally hyperbolic spacetimes. A globally hyperbolic
spacetime will be denoted by a pair $(M,g)$ where $M$ is the spacetime manifold
(assumed to be $C^\infty$) and $g$ is the Lorentzian metric. We will consider the
case of spacetime dimension equal to 4 with metric signature $(+ - - -)$, but
most of our considerations can be readily generalized, with appropriate modifications,
to arbitrary spacetime dimensions. We recall that global hyperbolicity means that
the spacetime is time-orientable and possesses Cauchy surfaces \cite{WaldGR,Baeretal}. 
Our conventions for curvature quantities, like in \cite{FewOst}, are those of 
Birrell and Davies, i.e. [-,-,-] in the classification scheme of Misner,
Thorne and Wheeler.

The classical linear scalar field $\varphi$ on a globally hyperbolic spacetime 
$(M,g)$ obeys the field equation
\begin{equation} \label{fieldeqn}
(\nabla^\mu\nabla_\mu + \xi R + m^2)\varphi = 0
\end{equation}
where $\nabla$ is the covariant derivative of $g$ and $R$ is the scalar curvature
corresponding to $g$; the constants $\xi \ge 0$ and $m \ge 0$ are the curvature coupling and
the mass parameters, respectively. The case $\xi = 0$ corresponds to minimal
coupling. 

The quantization of the system proceeds as follows. Owing to global hyperbolicity,
there are (for each fixed $\xi$ and $m$) two uniquely determined linear maps $E^\pm: C_0^\infty(M,\mathbb{R})
\to C^\infty(M,\mathbb{R})$ so that 
\begin{equation}
 E^\pm (\nabla^\mu\nabla_\mu + \xi R + m^2)f = f = (\nabla^\mu\nabla_\mu + \xi R + m^2)E^\pm f
\end{equation}
holds for all $f \in C_0^\infty(M,\mathbb{R})$, and additionally,
${\rm supp}(E^\pm f) \subset J^\pm({\rm supp}(f))$, where $J^\pm(G)$ is the
causal future/past set of $G \subset M$ \cite{Baeretal}. These are called the
advanced/retarded fundamental solutions of the wave-operator $(\nabla^\mu\nabla_\mu + \xi R + m^2)$, and with their help one can construct the real bilinear form
\begin{equation}
\mathscr{E}(f_1,f_2) = \int_M f_1(x) (E^- f_2 - E^+ f_2)(x)\, d{\rm vol}(x)
\end{equation} 
on $C_0^\infty(M,\mathbb{R})$ which turns out to be antisymmetric. Note that $\mathscr{E}$
is uniquely determined by $(M,g)$, $\xi$ and $m$.  
Fixing $\xi$ and $m$, one can now define the complex $*$-algebra 
$\mathcal{A}(M,g) = \mathcal{A}((M,g),\xi,m)$ with unit element $1$ as
being generated by a family of objects $\phi(f)$, $f \in C_0^\infty(M,\mathbb{R})$ which
are required to fulfil the following relations:
\\[2pt]
(a) $f \mapsto \phi(f)$ is real-linear,
\\[2pt]
(b) $\phi(f)^* = \phi(f)$,
\\[2pt]
(c) $\phi((\nabla^\mu\nabla_\mu + \xi R + m^2)f) = 0$,
\\[2pt]
(d) $[\phi(f_1),\phi(f_2)] = i\mathscr{E}(f_1,f_2) 1$.
\\[2pt]
Here $[A,B] = AB -BA$ denotes the commutator. Since the generators $\phi(f)$ of
$\mathcal{A}(M,g)$ obey, according to (d), the canonical commutation relations
in a covariant manner, one has thus obtained a quantization of the system in an
abstract form. The hermitean elements in $\mathcal{A}(M,g)$ correspond to observables of the quantized system, but they do not contain all observables that one may wish to consider,
so that the algebra $\mathcal{A}(M,g)$ will have to be enlarged to include those
additional observables as well. We will come back to this point. For the moment, a
{\em state} $\omega$ of the quantized linear scalar field is, by definition, a linear
functional $\omega: \mathcal{A}(M,g) \to \mathbb{C}$, $A \mapsto \omega(A) \equiv
\langle A \rangle_\omega$, with the additional property that $\omega$ is positive,
meaning $\omega(A^*A) \ge 0$ for all $A \in \mathcal{A}(M,g)$, and also with the property
that $\omega$ is normalized, i.e.\ $\omega(1) = 1$. Now it is known from examples that not every state according to this definition corresponds to a physically reasonable configuration
of the system and that selection criteria for physical states are needed. In the case of 
the linear fields on curved spacetime, the best candidates for physical states are
quasifree Hadamard states, and most other physical states can be derived from those \cite{VeHad}. 

We will very briefly summarize the concept of a quasifree Hadamard state. (For a more
in-depth discussion, see \cite{KayWald}.)
For any state $\omega$ on $\mathcal{A}$, the $n$-point functions are the maps
\begin{equation}
f_1 \otimes \cdots \otimes f_n \mapsto {\mathscr{W}}^\omega_n(f_1,\ldots,f_n) = \omega(\phi(f_1) \cdots \phi(f_n))\,.
\end{equation}
Clearly, each state on $\mathcal{A}(M,g)$ is determined by all the $n$-point functions. A quasifree state $\omega$ on
$\mathcal{A}(M,g)$ is a state which is entirely determined by its two-point function, by requiring
that the truncated $n$-point functions vanish \cite{BraRo}. This can also be expressed as
$\omega({\rm e}^{it\phi(f)}) = {\rm e}^{- t^2 \mathscr{W}^\omega_2(f,f)/2}$, to be interpreted as a sequence
of relations in the sense of formal power series in $t$. A (quasifree) state $\omega$ is called 
Hadamard state if its two-point function is of Hadamard form. This is the case, in turn, if
for any geodesic convex neighbourhood $N$ of any given point $x_o$, and any
time function $t$ on the underlying spacetime $M$,
one can find a sequence $H^{\omega}_k \in C^k(N\times N,\mathbb{C})$ such that for all
$f_1,f_2 \in C^\infty_0(N,\mathbb{R})$ one has 
\begin{eqnarray}
& & {\mathscr{W}^\omega_2 (f_1,f_2)} \\
& & = \ \lim_{\varepsilon \to 0+}\,\frac{1}{4 \pi^2} \int_{N \times N} (\, G_{k,\varepsilon}(x,x')
 + H_k(x,x')\, )\, f_1(x) f_2(x')\,d{\rm vol}(x) \, d{\rm vol}(x')\,, \nonumber
\end{eqnarray}
where 
\begin{eqnarray}
\label{HadParDefn}
{G_{k,\varepsilon}(x,x')} 
& = & \frac{U(x,x')}{\sigma(x,x') + 2i(t(x) - t(x'))\varepsilon + \varepsilon^2} \\
  & &   + \mathcal{V}_k (x,x'){\rm ln}(\sigma(x,x') + 2i(t(x) - t(x'))\varepsilon + \varepsilon^2)\,. \nonumber
\end{eqnarray}     
Here,
\begin{equation}
\label{HadVDefn}
\mathcal{V}_k(x,x') = \sum_{j= 0}^k U_j(x,x') \sigma(x,x')^j\,,
\end{equation}
$\sigma(x,x')$ is the squared geodesic distance from $x$ to $x'$\footnote{
Following \cite{FewSmith}, we choose $\sigma$ positive for $x$ and $x'$
spacelike related and negative for $x$ and $x'$ timelike related, so that e.g.
on Minkowski spacetime it is given by $\sigma(x,x')=-g_{ab} (x-x')^a (x-x')^b$},
and $U$ and $U_j$ 
are smooth functions on $N \times N$ determined by the Hadamard recursion relations.
Thus, the term $G_{k,\varepsilon}$ is, for each $k$, determined by the local
spacetime geometry and the parameters $\xi$ and $m$ of the scalar field equation.
For later use, we define the distribution
\begin{equation}
\mathscr{G}_k(f_1,f_2) = \lim_{\varepsilon \to 0+}\,\frac{1}{4 \p^2} \int
G_{k,\varepsilon}(x,x')\,f_1(x)f_2(x') \,d{\rm vol}(x)\,d{\rm vol}(x')
\end{equation}
for test-functions $f_1,f_2$ supported in a geodesic convex neighbourhood $N$.

It is worth noting that the existence of very many quasifree Hadamard states (spanning an
infinite dimensional space) has been established for the linear scalar field on all
globally hyperbolic spacetimes. Moreover, in stationary, globally hyperbolic spacetimes, the canonical
ground state as well as the thermal equilibrium states (KMS states) are known to
be quasifree Hadamard states \cite{SahlVe}. 

Let us now turn to the concept of local thermal equilibrium states introduced in \cite{BuOjiRoo} and further
investigated in \cite{Bu,BuSchl}. This will be done first for the case that the underlying spacetime is
just Minkowski spacetime. Our discussion here is limited to the linear scalar field, but as 
explained in \cite{BuOjiRoo}, the discussion can be generalized to include general quantum field theories.
For the quantized linear scalar field on Minkowski spacetime, there is in each Lorentz frame a 
unique quasifree thermal equilibrium state at given temperature. Actually, for fixed temperature this
state depends only on the time-direction of the Lorentz frame. Let $e_0$ be such a time-direction,
i.e.\ a timelike, future-pointing unit vector on Minkowski spacetime, and let $e_1,e_2,e_3$ be a
set of spacelike unit vectors so that $e = (e_0,e_1,e_2,e_3)$ forms an orthonormal tetrad on
Minkowski spacetime. When choosing coordinates $(x^0,x^1,x^2,x^3)$ on Minkowski spacetime such
that the coordinate axes are aligned with the tetrad, the two-point function $\mathscr{W}^{\beta e}_2$
of the unique quasifree thermal equilibrium (KMS) state $\omega^{\beta e}$ at inverse temperature $\beta > 0$
\footnote{$\beta = 1/(k_B {\tt T})$ where $k_B$ is Boltzmann's constant and {\tt T} is absolute temperature}
with respect to the Lorentz frame defined by $e$ is given by
\begin{equation} \label{equili}
{\mathscr{W}}^{\beta e}_2(x,x') = \int {\rm e}^{-i(x - x')^\mu p_\mu} \epsilon(p_0)\delta (p_\mu p^\mu - m^2)
\, \frac{d^4p}{(2\pi)^3(1- {\rm e}^{-\beta p_0})}\,,
\end{equation}
to be interpreted in the sense of distributions, where $\epsilon(p_0)$ is the sign function of $p_0$.
The uniqueness implies that every intensive thermal property (e.g.\ pressure, density etc) can be
expressed as a function of the timelike vector $\beta e_0$.

The passage from these global equilibrium states to states of local thermodynamic equilibrium now uses
spaces $S_x$ of observables located at spacetime {\em points} $x$. Mathematically this implies
that these observables are no longer defined as operators but only as quadratic forms (their
products are usually not defined). In physical terms, the observables in $S_x$ should model idealized
limits of measurements of intensive thermal properties of states in smaller and smaller spacetime regions.
This culminates in the requirement that, for $s(x) \in S_x$, the functions
$$ \Phi_s(x) = \omega^{\beta e}(s(x)) $$
have to be independent of $x$ and non-constant as functions of $\beta$. Furthermore, for many
$s(x) \in S_x$ one can identify the thermal quantity to which $\Phi_s$, as a function of $\beta$,
actually corresponds. As an example, one calculates that for $s(x) = : \phi^2:(x)$ (Wick-square) one
obtains for the massless case
$$ \Phi_{:\phi^2:} = \omega^{\beta e}(:\phi^2:(x)) = \frac{1}{12 \beta^2} = \frac{k_B^2 {\tt T}^2}{12}
  $$
which leads to the identification of this observable as a ``scalar thermometer'', giving the square of
the local temperature {\tt T} times some fixed constant. Choosing appropriate units to measure {\tt T},
this constant can be set equal to $1$, justifying the notation ${\tt T}^2$ for the expectation value
of the Wick-square in thermal equilibrium states already alluded to above. Actually this seems quite
similar to what one would do to construct a thermometer in the laboratory: Take some (small) device
which, when exposed to a situation known to be in equilibrium at some temperature {\tt T}, gives
a reading which is a simple function of {\tt T}.

In the investigations of local thermal equilibrium for linear scalar fields $\phi(x)$ on Minkowski spacetime
carried out in previous articles \cite{BuOjiRoo,Bu,BuSchl}, the spaces $S_x$ are chosen as the linear spaces generated by
elements $s(x) = \balder_{\boldsymbol{\mu}} :\phi^2:(x)$, referred to as the {\em balanced derivatives}
of the Wick-squared field $:\phi^2:(x)$.
\footnote{In the references \cite{BuOjiRoo,Bu}, the notation $\balder^{\boldsymbol{\mu}}$ is used, but we prefer to
view $\boldsymbol{\mu}$ as a co-tensor index}
 Here, $\boldsymbol{\mu} =(\mu_1,\ldots,\mu_n) \in \mathbb{N}^n$
is a multi-index of arbitrary length $n$, and the balanced derivatives are defined by
\begin{equation} \label{balanced}
\balder_{\boldsymbol{\mu}}:\phi^2:(x) = \lim_{\zeta \to 0}
\partial_{\boldsymbol{\mu}} \,\left(\phi(x + \zeta)\phi(x - \zeta) - \omega^{\rm vac}
(\phi(x + \zeta)\phi(x - \zeta))1 \right)
\end{equation}
where $\partial_{\boldsymbol{\mu}} = \partial_{\zeta^{\mu_1}} \cdots \partial_{\zeta^{\mu_n}}$, and where
$\omega^{\rm vac}$ is the vacuum state. The limit is taken along spacelike directions $\zeta$, so that
$\phi(x + \zeta)\phi(x -\zeta)$ is well defined as a quadratic form, and the limit defines an operator-valued
distribution after smearing in $x$ with test-functions. For multi-index length equal to 0, the balanced derivative 
equals just the Wick-square $:\phi^2:(x)$. For linear fields on Minkowski spacetime, this definition of
the Wick-square coincides with the usual normal ordering prescription. Owing to the translation invariance of
the KMS-states $\omega^{\beta e}$ one can easily check that the thermal functions
$\Phi_{\balder_{\boldsymbol{\mu}}:\phi^2:} = \omega^{\beta e}(\balder_{\boldsymbol{\mu}} : \phi^2 :(x))$ are
independent of $x$.

Following \cite{BuOjiRoo,Bu}, a state of the linear scalar field on Minkowski spacetime is said to be
{\em locally in thermal equilibrium} at a spacetime point $x$ if it looks like a global
thermal equilibrium state $\omega^{\beta e}$ as far as the expectation values of elements in
$S_x$ are concerned. The following definition, taken from \cite{BuOjiRoo}, expresses this more formally.
\begin{Def} \label{LTEmink}
A state $\omega$ of the quantized linear scalar field $\phi(x)$ on Minkowski spacetime is called
{\em $S_x$-thermal} at the spacetime point $x$ if there are an orthonormal tetrad $e$ with
$e_0$ timelike and future-pointing, and $\beta > 0$, such that 
\begin{equation} \label{ltex}
 \omega(s(x)) = \omega^{\beta e}(s(x))
\end{equation}
holds for all $s(x) \in S_x$, where $S_x$ is spanned by $\balder_{\boldsymbol{\mu}} :\phi^2:(x)$ as
$\boldsymbol{\mu}$ ranges over all multi-indices. 
\\[6pt]
If $\mathcal{O}$ is some open set of spacetime points, a state $\omega$ of the quantized linear
scalar field on Minkowski spacetime is called {$S_{\mathcal O}$-thermal} if \eqref{ltex} holds and
$\omega(s(x))$ varies continuously with $x$, for $s(x) \in S_x$ and $x \in \mathcal{O}$. That means,
$\omega$ is $S_x$-thermal at each $x \in \mathcal{O}$, where $\beta$ and $e$ in \eqref{ltex} may
vary with $x$.
\end{Def}
For $S_{\mathcal{O}}$-thermal states, the expectation value $\langle \Phi_s\rangle_\omega(x)$ of
an extensive thermal quantity $\Phi_s$ at $x$, whose local measurement is modelled by $s(x) \in S_x$,
is then given as $\langle \Phi_s\rangle_\omega(x) = \omega(s(x))$.
This leads for $S_{\mathcal{O}}$-thermal states to an assignment of thermal quantities to each
$x \in \mathcal{O}$ whose values in general vary with $x$, and this assignment is consistent in
the sense that relations among the thermal quantities (like equations of state) also hold
at each point. In the case of $s(x) = :\phi^2:(x)$, one obtains in this way for an $S_{\mathcal{O}}$-thermal
state $\omega$ an assignment of the expected squared temperature to each spacetime point $x \in O$.

The concept of states which are locally in thermal equilibrium has also been generalized in
\cite{BuOjiRoo} to allow mixtures of global thermal equilibrium states on the right hand side of 
\eqref{ltex}. We will summarize this generalized concept in Sec.\ 5.

In attempting to extend the concept of local thermal equilibrium states to quantum fields in curved
spacetime, one faces a couple of difficulties which are, of course, connected to the occurrence
of curvature and the related lack of global vacuum states and global equilibrium states. Primarily,
these difficulties are:
\par \noindent
(i) The definition \eqref{balanced} of balanced derivatives $\balder_{\boldsymbol{\mu}}:\phi^2:(x)$
uses the Minkowski vacuum state $\omega^{\rm vac}$ as preferred vacuum state.
\par \noindent
(ii) Moreover, the definition \eqref{balanced} uses the affine space structure of Minkowski spacetime.
\par \noindent
(iii) Def.\ \ref{LTEmink} uses global thermal equilibrium states $\omega^{\beta e}$ on Minkowski spacetime,
for which there is no counterpart on generic curved spacetimes.
\par \noindent
Thus, there is no verbatim translation of the concept of local thermal equilibrium states given in 
Def.\ \ref{LTEmink}.

It is clear that problems (i) and (ii) concern the definition of balanced derivatives of a Wick-squared
quantum field in curved spacetime. We will soon turn to that problem.
Assuming that the definition of balanced derivatives in curved spacetime is settled, a proposal was
made in \cite{BuSchl} to surpass problem (iii). The idea is to define that a state $\omega$ of the 
quantized linear scalar field $\phi$ on a curved spacetime $(M,g)$ is $S_x$-thermal at a point
$x$ in $M$ if $\omega(s(x)) = \omM^{\beta e}(\sM(\xM))$
holds for all $s(x) \in S_x$. Here, $\omM^{\beta e}$ is a thermal equilibrium state
of the free scalar field $\phM$ (with same parameters as $\phi$) on Minkowski spacetime
$\MM$, $\xM$ is a point in $\MM$, and $\sM(\xM)$
is the flat space counterpart of $s(x)$. To explain what this latter phrase means precisely can be seen as part of
the definition of balanced derivatives in curved spacetime, but certainly one would require that
$\sM(\xM)$ corresponds to a balanced derivative of the Wick-square of $\phM$
if $s(x)$ corresponds to a balanced derivative of the Wick-square of $\phi$.
In the approach of \cite{BuSchl}, the requirement of local thermality on $\omega$ is thus not implemented
by comparing expectations values of pointlike thermal observables with the corresponding expectation
values in a global thermal state (as such states need not exist), but with the ``flat space version'' of
a thermal equilibrium situation for the quantum field. The motivation for this approach is that
$S_x$-thermality is a pointwise property which should not be affected by curvature; this, in turn, rests
largely on the equivalence principle.

For a linear scalar quantum field $\phi$ on a curved spacetime it is simple enough to know what
its flat space counterpart $\phM$ should be. However, as pointed out in \cite{BuSchl},
one may invoke the concept of a local covariant quantum field theory \cite{BFV,HolWald-wick1} to know this also for more
general types of quantum fields. The concept of local covariance affects also the elements 
$s(x) \in S_x$; in our situation where we start from a linear scalar field $\phi$ in curved
spacetime -- which is known to have the structure of a local covariant quantum field -- the
balanced derivatives of Wick-squares of $\phi$ should be defined in such a way that they are also
local covariant quantum fields.

Let us thus discuss our proposal for the generalization of the concept of balanced derivatives in a curved
spacetime. Our discussion is greatly facilitated by the circumstance that for the purpose of deriving QEIs
for LTE states we need only focus on balanced derivatives up to second order, corresponding to a multi-index
length of $\boldsymbol{\mu}$ not greater than two. Accordingly, we will define the LTE property on curved spacetime
only with balanced derivatives of the Wick-squared field up to second order, see below.

We proceed in two steps. First, we shall consider the generalization of expressions like $\partial_{\zeta^\mu}\partial_{\zeta^\nu}
f(x + \zeta,x - \zeta)|_{\zeta = 0}$ for $C^2$-functions $f$ from Minkowski spacetime to curved spacetime. This discussion
is entirely of differential geometric nature. In a second step, we have to give a generalization of the quantity
$$ f(x + \zeta,x - \zeta) = \omega(\phi(x + \zeta)\phi(x -\zeta)) - \omega^{\rm vac}(\phi(x + \zeta)\phi(x - \zeta)) $$
for Hadamard states $\omega$ on curved spacetime where there is no counterpart of $\omega^{\rm vac}$. In doing this we have
to ensure, as mentioned, that the resulting balanced derivatives of the Wick-ordered linear scalar field give rise to
local covariant quantum fields.

Turning to the first step, let $(M,g)$ be a spacetime and suppose that $N$ is a geodesically convex neighbourhood
of some point $x$ in $M$. The exponential map at $x$ will be denoted by ${\rm exp}_x$. A fairly obvious generalization
of the first balanced derivative of a function $f \in C^2(N \times N)$ arises by requiring
$$ v^a \balder_af(x) = \left. \frac{d}{d\lambda} \right|_{\lambda = 0} f({\rm exp}_x(\lambda v),{\rm exp}_x(-\lambda v))$$
for all spacelike vectors $v = v^a \in T_xM$ lying in ${\rm exp}_x^{-1}(N)$. By linearity, this determines a co-vector
$\balder_a f(x)$ in $T^*_xM$.
We define $T(^{r\,r'}_{s\,s'})(M \times M)$ as the bundle over $M \times M$ whose fibre at $(y,y') \in M \times M$ is
given by $T_y(^r_s)M \otimes T_{y'}(^{r'}_{s'})M$, where $T_y(^r_s)M$ coincides with the space of $r$-fold contravariant
and $s$-fold covariant tensors at $y$. If $(y,y') \mapsto V(y,y')$ is any $C^1$ section in $T(^{r\,r'}_{s\,s'})(M \times M)$,
then we denote by $\nabla_a V$  the covariant derivative with respect to the $y$-entry and by $\nabla_{a'}V$ the
covariant derivative with respect to the $y'$-entry. Furthermore, we denote by
$V\lfloor_x = V(x,x)$ the coincidence value of $V(y,y')$ for $y=x=y'$.     
With these conventions, one has
$$ \balder_af(x) = \nabla_a f \lfloor_x - \nabla_{a'} f\lfloor_x \,,$$
and if $f$ is $C^2$, this defines a $C^1$ co-vector field as $x$ varies.

The second order balanced derivative $\balder_{ab}f(x)$ can then be defined as follows. $\nabla_a f(y,y')$
is a $y'$-dependent co-vector at $y$, and vice versa for $\nabla_{a'} f(y,y')$. Let $v = v^a$ be a (spacelike)
vector in $T_xM$ which lies in ${\rm exp}_x^{-1}(N)$, so that $\eta_v : \lambda \mapsto {\rm exp}_x(\lambda v)$
$( -1 \le \lambda \le 1)$ is the geodesic determined by $v$ at $x$. Correspondingly, we can define the map of
parallel transport $P_{v,\lambda} : T^*_{{\rm exp}_x(\lambda v)}M \to T^*_xM$ of co-vectors from
${\rm exp}_{\lambda v} = \eta_v(\lambda)$ to $x = \eta_v(0)$ along the geodesic $\eta_v$. A geometrically natural
definition of the second order balanced derivative $\balder_{ab}f(x)$ of $f$ at $x$ is then obtained by demanding
that
\begin{eqnarray*}
 v^a w^b \balder_{ab}f(x) & = & \left. \frac{d}{d \lambda} \right|_{\lambda = 0}
                               w^b P_{v,\lambda}\nabla_b f({\rm exp}_x(\lambda v),{\rm exp}_x(-\lambda v)) \\ 
 & & - \left. \frac{d}{d \lambda} \right|_{\lambda = 0}
                               w^{b'} P_{-v,\lambda}\nabla_{b'} f({\rm exp}_x(\lambda v),{\rm exp}_x(-\lambda v)) 
\end{eqnarray*}
holds for all (spacelike) vectors $v,w \in T_xM$ with $v \in {\rm exp}_x^{-1}(N)$. Using the properties of the parallel
transport, it follows that 
$M \owns x \mapsto \balder_{ab}f(x)$ is a continuous (if $f$ is $C^2$) $(^0_2)$-tensor field on $M$, and
\begin{eqnarray}
\label{BalDerDiffGeo}
\balder_{ab}f(x) & = & \nabla_a \nabla_b f\lfloor_x - \nabla_a \nabla_{b'} f \lfloor_x \\
                  & & - \nabla_{a'} \nabla_{b} f \lfloor_x + \nabla_{a'} \nabla_{b'} f \lfloor_x \,. \nonumber 
\end{eqnarray}                                
Note here that the covariant derivatives on the right hand side act on $y$ for unprimed indices and
on $y'$ for primed indices, and primed and unprimed tensor indices are identified at the coincidence
point $y = x = x'$.

Turning to the second step, suppose that $(M,g)$ is a globally hyperbolic spacetime, and that  
$N$ is a geodesic convex neighbourhood of some point $x \in M$. Then define the distributions $(k \ge 2)$
$$ \widetilde{\mathscr{G}}_k(f_1,f_2) = \frac{1}{2} (\mathscr{G}_k(f_1,f_2) + \mathscr{G}_k(f_2,f_1)) + i \mathscr{E}(f_1,f_2)\,,
\quad f_1,f_2 \in C_0^\infty(N,\mathbb{R})\,,$$
where $\mathscr{G}_k$ and $\mathscr{E}$ are defined above for the quantized linear scalar field $\phi$ on
$(M,g)$.

Next, let $\omega$ be a quasifree Hadamard state of $\phi$ on $(M,g)$; then define the point-split renormalized
two-point function obtained by subtracting the symmetrized Hadamard parametrix (SHP) $\widetilde{\mathscr{G}}_k$ from the 
two-point function:
$$ \mathscr{W}^{\scSHP}_{\omega,k}(f_1,f_2) = \mathscr{W}_2^\omega(f_1,f_2) - \widetilde{\mathscr{G}}_k(f_1,f_2)\,,
         \quad f_1,f_2 \in C_0^\infty(N,\mathbb{R})\,.$$
A first observation is that $\mathscr{W}^{\scSHP}_{\omega,k}$ is symmetric, $\mathscr{W}^{\scSHP}_{\omega,k}(f_1,f_2)
= \mathscr{W}_{\omega,k}^{\scSHP}(f_2,f_2)$. Furthermore, for $k \ge 2$, $\mathscr{W}_{\omega,k}^{\scSHP}$ is given
$$
 \mathscr{W}_{\omega,k}^{\scSHP}(f_1,f_2) = \int_{N \times N} W_{\omega,k}^{\scSHP}(y,y') f_1(y) f_2(y') \,d{\rm vol}(y)
 \,d{\rm vol}(y')\,, \quad f_1,f_2 \in C_0^\infty(N,\mathbb{R})\,,$$
 with $W_{\omega,k}^{\scSHP} \in C^2(N \times N)$ for $k \ge 2$, which follows from the definition of Hadamard
 form (see also the arguments in \cite{FewSmith}).

Consequently, one may now define for any quasifree Hadamard state $\omega$ the expectation value of the SHP-Wick square of
$\phi$ and the corresponding second balanced derivatives in the following way:
\begin{Def}
 \begin{eqnarray} \label{expomega}
 \omega(:\phi^2:_{\scSHP}(x)) & = & W^{\scSHP}_{\omega,k} \lfloor_x\,, \\
 \label{expomegaii}
 \omega(\balder_{ab} :\phi^2:_{\scSHP}(x)) & = & \balder_{ab}W^{\scSHP}_{\omega,k} \lfloor_x
 \end{eqnarray}
for $x \in M$, with $k \ge 2$.
\end{Def}
We add a few observations to this definition.
\\[6pt]
($\alpha$) One can likewise define the first balanced derivative of $\omega(:\phi^2:_{\scSHP}(x))$, but since 
$W^{\scSHP}_{\omega,k}$ is symmetric, its first balanced derivative vanishes. This is similar to the
property of balanced derivatives of the Wick-square of the quantized linear scalar field $\phM$
on Minkowski spacetime, which can be traced back to the symmetry of 
$$\omM(\phM(y)\phM(y')) - \omega^{\rm vac}(\phM(y)\phM(y'))$$
with respect to $y$ and $y'$, for each quasifree Hadamard state $\omM$ of $\phM$. This
provides motivation why we define the Wick-ordering by subtraction of the {\em symmetrized} Hadamard parametrix.
\\[6pt]
($\beta$) Actually, $W_{\omega,k}^{\scSHP}(y,y')$ and $\nabla_a \nabla_{b'} W_{\omega,k}^{\scSHP}(y,y')$ depend
on the time-function $t$ entering into the definition of $G_{k,\epsilon}$, but for $k \ge 2$, this dependence
vanishes in the coincidence limit $y = x = y'$. Similarly, $W_{\omega,k}^{\scSHP}\lfloor_x$ and $\nabla_a \nabla_{b'}
W_{\omega,k}^{\scSHP}\lfloor_x$ are independent of $k$ for $k \ge 2$. 
\\[6pt]
($\gamma$) 
One purpose of using the point-split renormalization by subtraction of the symmetrized Hadamard parametrix 
$\widetilde{\mathscr{G}}_k$ is that the latter is a locally constructed geometric quantity which is state
independent, so that $:\phi^2:{}_{\scSHP}$ and $\balder_{ab}:\phi^2:{}_{\scSHP}$ become local, covariant fields.
If one applies this technique to the linear scalar field $\phM$ on Minkowski spacetime, one finds that
$:\phM^2:{}_{\scSHP}$ and $\balder_{\mu\nu}:\phM^2:{}_{\scSHP}$ deviate for
$m > 0$ from the usual flat-space definitions of $:\phM^2:$ and $\balder_{\mu\nu}:\phM^2:$,
described above, by constants. Concretely, using the expression for
$\mathscr{W}_2^{{\omega}{}^{\rm vac}} - \widetilde{\mathscr{G}}_k$ ($k \ge 2$),
one can calculate that, for each Hadamard state $\omM$ of $\phM$, one has
\begin{eqnarray}
\omM(:\phM^2:{}_{\scSHP}(\xM)) & = & \omM(:\phM^2:(\xM))
     + c_{0,m} \,, \label{Wickdiff1} \\
\omM(\balder_{\mu\nu}:\phM^2:{}_{\scSHP}(\xM)) & = & \omM(\balder_{\mu\nu}:\phM^2:(\xM))
     + c_{2,m}\eta_{\mu\nu}\,, \label{Wickdiff2}
\end{eqnarray} 
at all points in Minkowski spacetime (details in Appendix B). Here, $m$ is the mass parameter of the linear scalar field, and $\eta_{\mu\nu}$
is the Minkowski metric. The constants $c_{0,m}$ and $c_{2,m}$ vanish for $m = 0$; for $m > 0$, they are given by
\begin{eqnarray}
c_{0,m} &= & \frac{m^2}{(4\pi)^2}\left[\ln \left(\frac{{\rm e}^{2\gamma} m^2}{4}\right) -1 \right] \,,\\
c_{2,m} &= & -\frac{m^4}{(4\pi)^2}\left[\ln \left(\frac{{\rm e}^{2\gamma} m^2}{4}\right) - \frac{5}{2} \right] \,,
\end{eqnarray}
where $\gamma$ denotes the Euler-Mascheroni constant. This needs to be taken into account in the definition of thermal
equilibrium states below.
\\[6pt]
We can now define the concept of a local thermal equilibrium state on a globally hyperbolic curved spacetime
$(M,g)$. Let $e = (e_0,e_1,e_2,e_3)$ be an orthonormal tetrad at $x \in M$, with $e_0$ timelike
and future-pointing. Then $e$ induces an identification of $T_xM$ with Minkowski spacetime $\MM$,
whereupon $e$ is identified with a basis of $\MM$, again with $e_0$ timelike and future-pointing
in Minkowski spacetime. This identification is used in the following definition.
\begin{Def} \label{Def-LTE-CST}
 Let $\omega$ be a state with two-point function of Hadamard form for the quantized linear scalar field
$\phi$ on a globally hyperbolic spacetime $(M,g)$. Then let $\phM$ denote the quantized linear scalar
field, with the same parameters as $\phi$, on Minkowski spacetime $\MM$.
\\[6pt]
(a) \quad We say that $\omega$ is $S_x^{(2)}$-{\em thermal} at a point $x \in M$ if, with some orthonormal
tetrad $e=(e_0,e_1,e_2,e_3)$ at $x$ such that $e_0$ is timelike and future-pointing, there is a thermal
equilibrium state $\omM^{\beta e}$ of $\phM$ so that -- upon identification
of $e$ with a basis tetrad of $\MM$ -- the equalities
\begin{eqnarray} 
\omega(:\phi^2:{}_{\scSHP} (x)) & = & \omM^{\beta e}(:\phM^2:{}_{\scSHP}(\xM))\label{beins} \\
                          & = & \omM^{\beta e}(:\phM^2:(\xM)) + c_{0,m}\,, \nonumber \\
v^aw^b\omega(\balder_{ab}:\phi^2:{}_{\scSHP} (x)) & = & 
v^{\mu}w^{\nu}\omM^{\beta e}(\balder_{\mu\nu}:\phM^2:{}_{\scSHP}(\xM)) \label{bzwei} \\
& = & v^\mu w^\nu \omM^{\beta e}(\balder_{\mu\nu}:\phM^2:(\xM)) + c_{2,m} v^\mu w^\nu \eta_{\mu\nu} \nonumber
\end{eqnarray}
hold for all (spacelike) vectors $v,w \in T_xM$ with coordinates $v^\mu e_\mu = v$, $w^\nu e_\nu = w$, for
some $\xM \in \MM$. (By translation-invariance of $\omM^{\beta e}$, the 
particular choice of $\xM$ is irrelevant.)
\\
(b) \quad Let $N$ be a subset of $M$. We say that $\omega$ is $S_{N}^{(2)}$-{\em thermal}
if \\
 $\omega(:\phi^2:{}_{\scSHP}(x))$ and $\omega(\balder_{ab}:\phi^2:{}_{\scSHP}(x))$ are continuous in $x \in 
N$ and if, for each $x \in N$, $\omega$ is $S_x^{(2)}$-thermal at $x$.
\end{Def}
The definition of $S_x^{(2)}$-thermal states demands the coincidence of expectation values of the SHP Wick square
of $\phi$ and its balanced derivatives up to second order with the thermal equilibrium situation in flat spacetime.
This amounts to saying that $S_x^{(2)}$ consists of linear combinations of the unit operator $1$ and of the
quadratic forms $:\phi^2:{}_{\scSHP}(x)$ and $\balder_{ab}:\phi^2:{}_{\scSHP}(x)$ whose evaluations (i.e.\ expectation
values) on states $\omega$ are given by \eqref{expomega} and \eqref{expomegaii}. Thus, for the linear
scalar field on Minkowski spacetime, $S^{(2)}_x$ is a small subset of $S_x$, and thus an $S_x^{(2)}$-thermal state
fulfills less constraints on its thermal properties than an $S_x$-thermal state. We shall not follow up that matter
at this point. Our definition of $S_x^{(2)}$-thermal states (or $S_{\mathcal{O}}^{(2)}$-thermal states) turns out
to be sufficient to derive quantum energy inequalities.

Given an $S_x^{(2)}$-thermal state $\omega$, we shall now use the abbreviations
\begin{eqnarray}
 \vartheta^{\omega}(x) & =& \omega(:\phi^2:{}_{\scSHP}(x))\,, \\
 \varepsilon^{\omega}_{ab}(x) & = & -\frac{1}{4}\omega(\balder_{ab}:\phi^2:{}_{\scSHP}(x))\,.
\end{eqnarray}
Then we have
\begin{Lem} \label{Lemma}
Let $\phi$ be a linear scalar field on $(M,g)$, with mass parameter $m$,
 and let $\omega$ be an $S_{x}^{(2)}$-thermal state
$(x \in M)$, satisfying \eqref{beins} and \eqref{bzwei} for some $\beta >0$ and
an orthonormal tetrad $e = (e_0,e_1,e_2,e_3)$ at $x$. Then the following statements hold.
\\[6pt] 
(a) \quad $\vartheta^\omega(x) = \frac{1}{\beta^2}\chi_{0,m}(\beta) + c_{0,m}$,\ \ with
$$ \chi_{0,m}(\beta) = \frac{1}{2\pi^2} \int_0^\infty \frac{\rho^2 \ \ d\rho}
{({\rm e}^{\sqrt{\rho^2 +\beta^2m^2}} -1)\sqrt{\rho^2 + \beta^2m^2}}\,.$$
(b) \quad $\varepsilon^\omega_a{}^a(x) = m^2\chi_{0,m}(\beta) - c_{2,m}$
\\[6pt]
(c) \quad Suppose that $v$ is a lightlike vector at $x$, $v_av^a =0$, or a timelike vector at $x$ with unit proper length,
$v_av^a = 1$, and set $v^0 = (e_0)_av^a$. Then one has the bound
\begin{equation} \label{dd}
\zeta(4)\frac{6 (v^0)^2}{\pi^2 \beta^4} - v_av^a\frac{c_{2,m}}{4} \ge
 v^av^b\varepsilon_{ab}(x) \ge \frac{(v^0)^2}{\beta^4}\chi_{2,m}(\beta) - v_av^a\frac{c_{2,m}}{4}
\end{equation} 
where
\begin{equation}
\chi_{2,m}(\beta) = \frac{1}{2\pi^2} \int_0^\infty 
\frac{\rho^2 \sqrt{\rho^2 + \beta^2m^2}}
{{\rm e}^{\sqrt{\rho^2 + \beta^2m^2}} -1} \ d\rho
\end{equation}
($\zeta(4)$ is the value of the $\zeta$-function at $4$.)
\end{Lem}
\noindent
{\it Proof.} The proof is based on the fact that, with respect to coordinates induced by the basis tetrad $e$,
\begin{eqnarray*}
\omM^{\beta e}(:\phM^2:(\xM)) & = &
\frac{1}{(2\pi)^3}\int_{\mathbb{R}^3} \frac{1}{({\rm e}^{\beta p_0} - 1)p_0}\,d^3\underline{p}\,, \\
-\frac{1}{4}\balder_{\mu\nu}\omM^{\beta e}(:\phM^2:(\xM) ) & = &
\frac{1}{(2\pi)^3}\int_{\mathbb{R}^3} \frac{p_\mu p_\nu}{({\rm e}^{\beta p_0} - 1)p_0}\,d^3\underline{p}\,, \\
\end{eqnarray*}
where $(p_\mu)_{\mu =0,\ldots,3} = (p_0,\underline{p})$ and $p_0 = \sqrt{|\underline{p}|^2 + m^2}$ in the 
integrals. The stated relations then basically result from transforming the integrals into spherical
polar coordinates. For the upper bound in \eqref{dd} notice that the integrand is given by
$$ \left. (v^0\sqrt{|\underline{p}|^2 + m^2} - \underline{v} \cdot \underline{p})^2 \right/\left(({\rm e}^{\beta \sqrt{|\underline{p}|^2 + m^2}} -1)
\sqrt{|\underline{p}|^2 + m^2} \, \right)$$
which is bounded above by
$$ |\underline{p}|\left. \left(v^0  - \frac{\underline{v} \cdot \underline{p}}{\sqrt{|\underline{p}|^2 + m^2}}\right)^2\right/(\rm{e}^{\beta |\underline{p}|} -1)
 \,.$$
Upon integration over $\underline{p}$, this can be bounded by the integrand $2(v^0)^2|\underline{p}|/({\rm e}^{\beta
|\underline{p}|} -1)$, using that $(v^0)^2 - |\underline{v}|^2 = 1$ in the timelike case
and $(v^0)^2 - |\underline{v}|^2 = 0$ in the lightlike case. \hfill $\Box$
\\[10pt]
Now we introduce a set of states whose local temperature is bounded above by some fixed value.
\begin{Def} Let $\beta' > 0$, $x \in M$. Then we define $\mathcal{L}_{\beta'}(x)$ as the set of
all $S_x^{(2)}$-thermal states $\omega$ of the linear scalar field on $(M,g)$ so that
\begin{equation} \label{ltx}
\vartheta^\omega(x) \le \frac{1}{(\beta')^2}\chi_{0,m}(\beta') + c_{0,m}\,.
\end{equation}
If $N \subset M$, we define $\mathcal{L}_{\beta'}(N)$ as the set
of all $S^{(2)}_N$-thermal states of the linear scalar field on $(M,g)$ so that \eqref{ltx}
is fulfilled for all $x \in N$.
\end{Def}
In other words, $\omega$ is in $\mathcal{L}_{\beta'}(x)$ if the relations \eqref{beins} and \eqref{bzwei}
are fulfilled for $1/\beta < 1/\beta'$.

Now let $N$ be an open subset of $M$, and let $\gamma : [\tau_0,\tau_1] \to
N$, $\tau \mapsto \gamma(\tau)$ be a geodesic with affine parameter $\tau$, and
denote by $v^a = \dot{\gamma}{}^a$ the tangent vector field of $\gamma$. By the geodesic
equation, it holds that
$$ (v^av^b \nabla_a \nabla_b \vartheta^\omega)(\gamma(\tau)) =
 \frac{d^2}{d \tau^2} \vartheta^\omega(\gamma(\tau))\,.$$
Consequently, we obtain for $\omega \in \mathcal{L}_{\beta'}(N)$ and
$\eta \in C_0^2((\tau_0,\tau_1))$,
\begin{eqnarray} \label{estimate}
|\,\int \eta(\tau)(v^av^b \nabla_a \nabla_b \vartheta^\omega)(\gamma(\tau))\,d\tau\,|
& = & |\, \int \eta''(\tau) \vartheta^\omega(\gamma(\tau))\,d\tau\,| \\
& \le & ||\eta''||_{L^1}\left|\frac{1}{(\beta')^2}\chi_{0,m}(\beta') + c_{0,m}\right|\,. \nonumber
\end{eqnarray}
Here, $\eta''$ is the second derivative of $\eta$.

\section{Quantum Energy Inequalities}

Let $(M,g)$ be a globally hyperbolic spacetime, and let $\varphi$ be the classical linear scalar field on $(M,g)$
with mass parameter $m \ge 0$ and conformal coupling parameter $\xi$. If $\varphi$ is a field configuration,
i.e.\ a smooth solution to the field equation \eqref{fieldeqn}, then the corresponding classical stress-energy tensor
is a $(^0_2)$ co-tensor field $T^{(\varphi)}_{ab}$ given by 
\begin{eqnarray*}
T^{(\varphi)}_{ab}(x) & = & (\nabla_a\varphi(x))(\nabla_b\varphi(x)) + \frac{1}{2}g_{ab}(x)(m^2\varphi^2(x) -
 (\nabla^c\varphi)(\nabla_c\varphi)(x)) \\
 & & + \xi (g_{ab}(x) \nabla^c\nabla_c -\nabla_a\nabla_b - G_{ab}(x))\varphi^2(x)\,, \quad x \in M\,,
\end{eqnarray*}
where $G_{ab} = R_{ab} - \frac{1}{2}g_{ab}R$ is the Einstein tensor.

Now let $\phi$ be the quantized linear scalar field on $(M,g)$, corresponding to the choice of parameters 
$m$ and $\xi$. The definition of the renormalized expectation value of products and derivatives for the quantized linear scalar field $\phi$
in a state $\omega$ having two-point function of Hadamard form proceeds, similarly to what was done in
the previous chapter, by point-splitting and subtraction of the SHP (see \cite{Wald78, WaldQFT}). To this end,
we define:
\begin{eqnarray*}
 \omega(:\phi \nabla_a \phi:{}_{\scSHP}(x)) & = & \nabla_{a'}W^{\scSHP}_{\omega,k} \lfloor_x \\
 \omega(:\phi \nabla_a \nabla_b \phi:{}_{\scSHP}(x)) & = & \nabla_{a'}\nabla_{b'}{W}^{\scSHP}_{\omega,k}\lfloor_x \\
 \omega(:(\nabla_a \phi)(\nabla_b \phi):_{\scSHP}(x) & = & \nabla_a \nabla_{b'}{W}^{\scSHP}_{\omega,k}\lfloor_x
\end{eqnarray*} 
with $k \ge 2$, $x \in M$. (Note again that $a$ and $a'$ are identified upon taking the coincidence limit $y = x = y'$ on the right hand side of each equation.) Owing to the symmetry of $W^{\scSHP}_{\omega,k}$ ($k \ge 2$), one can easily check that the
following Leibniz rule is fulfilled for SHP Wick-products involving derivatives:
\begin{eqnarray*}
 \omega(\nabla_a(:\phi^2:{}_{\scSHP}(x))) & = & 2 \omega(:\phi \nabla_a \phi :{}_{\scSHP}(x))\,, \\
 \omega(\nabla_a(:\phi \nabla_b \phi :{}_{\scSHP}(x))) & = & \omega(:(\nabla_a \phi)(\nabla_b \phi):{}_{\scSHP}(x))
       + \omega(:\phi \nabla_a \nabla_b \phi:{}_{\scSHP}(x))\,.
\end{eqnarray*}
The renormalized expectation value of stress-energy is then obtained via replacing the classical expressions
$\varphi^2(x)$, $(\nabla_a\varphi(x))(\nabla_b\varphi(x))$, and so on, by $\omega(:\phi^2:_{\scSHP}(x))$, $\omega(:(\nabla_a\phi)(\nabla_b \phi):{}_{\scSHP}(x))$,
etc. Using also the Leibniz rule for SHP Wick products, this leads to
\begin{eqnarray*}
 & & \omega(T^{\scSHP}_{ab}(x)) = -\omega(:\phi \nabla_a \nabla_b \phi:{}_{\scSHP}(x) + \frac{1}{4}\nabla_a\nabla_b
:\phi^2:{}_{\scSHP}(x)) \\
       & & \quad \quad + 
           \left( \frac{1}{4} - \xi  \right)\,\omega\left(\nabla_a\nabla_b :\phi^2:{}_{\scSHP}(x)
  - g_{ab}(x)\nabla^c\nabla_c :\phi^2:{}_{\scSHP}(x)  \right) \\
  & & \quad \quad + \frac{1}{2} g_{ab}(x) \omega\left( :\phi\nabla^c\nabla_c\phi:{}_{\scSHP}(x) + m^2:\phi^2:{}_{\scSHP}(x) \right)\\
   & &  \quad \quad  - \xi G_{ab}(x)\omega(:\phi^2:{}_{\scSHP}(x))
\end{eqnarray*}
This expression, however, has the defect of a non-vanishing divergence. The way to cope with this problem,
following Wald \cite{Wald78,WaldQFT}, is like this: It can be shown that $\nabla^a\omega(T^{\scSHP}_{ab}(x)) = \nabla_b Q(x)$,
where (apart from a free constant which can be set to a preferred value depending on the mass parameter $m$) 
$Q$ is a function which is determined by the local geometry of $(M,g)$; in particular, $Q$ is independent of
the state $\omega$. One may therefore subtract the term $Q(x) g_{ab}(x)$ from $\omega(T^{\scSHP}_{ab}(x))$ 
to make the resulting quantity have vanishing divergence. There remains an ambiguity in that one may still add
other $(^0_2)$ co-tensor fields $C_{ab}$ which are determined by the local geometry of $(M,g)$ and have
vanishing divergence. We take here the same view as put forward in \cite{FewSmith}, namely that the specification
of $C_{ab}$ is a further datum of the underlying quantum field $\phi$ on $(M,g)$, in addition to
the parameters $m$ and $\xi$. An alternative, elegant method has been proposed by  Moretti \cite{Moretti},
which nevertheless we won't follow here mainly because we would like to maintain close contact to other works on
quantum energy inequalities. This understood, we finally define the renormalized expectation value of the stress
energy tensor in some state $\omega$ (with two-point function of Hadamard form) of the linear scalar field
$\phi$ on $(M,g)$ as
\begin{equation}
 \omega(T^{ren}_{ab}(x)) = \omega(T^{\scSHP}_{ab}(x)) - Q(x)g_{ab}(x) + C_{ab}(x)\,, \quad x \in M\,.
\label{RenEitExp}
\end{equation}
Again note that $Q$ and $C_{ab}$ are state-independent and constructed locally out of the spacetime metric
$g = g_{ab}$.    

Let us next observe that, for each state $\omega$ of $\phi$ with two-point function of
Hadamard form, and with $R = R^a_a$ denoting the scalar curvature, 
\begin{eqnarray*}
F(x) & = & \omega(:\phi(\nabla^a \nabla_a + m^2 + \xi R)\phi:{}_{\scSHP}(x)) \\
  & = & \omega(:\phi(\nabla^a\nabla_a \phi):{}_{\scSHP}(x)) +
      (m^2 + \xi R(x))\omega(:\phi^2:{}_{\scSHP}(x))
\end{eqnarray*}
is a continuous function of  $x \in M$, independent of the state $\omega$, entirely determined by the
local geometry of $(M,g)$ and the parameters $m$ and $\xi$ of $\phi$.
To see this, note that
$$ \omega(:\phi(\nabla^a \nabla_a + m^2 + \xi R)\phi:_{\scSHP}(x)) = (\nabla^{a'}\nabla_{a'} + m^2 + \xi R)
W^{\scSHP}_{\omega,k}\lfloor_x\,.
$$
On the other hand, $W^{\scSHP}_{\omega,k}(y,y')$ is the integral kernel of 
$\mathscr{W}^{\scSHP}_{\omega,k} =
\mathscr{W}^{\omega}_2 -
\widetilde{\mathscr{G}}_k$ ($k \ge 2$), and since 
$\mathscr{W}^{\omega}_2(f,(\nabla^b\nabla_b + m^2 + \xi R)h) = 0$, it follows that 
$(\nabla^{a'}\nabla_{a'} + m^2 + \xi R)W^{\scSHP}_{\omega,k}$ is independent of $\omega$ as
the $\omega$-dependent part of $W^{\scSHP}_{\omega,k}$ is annihilated by the
wave-operator $(\nabla^{a'}\nabla_{a'} + m^2 +\xi R)$. In consequence, $F(x) = 
(\nabla^{a'}\nabla_{a'} + m^2 + \xi R)W^{\scSHP}_{\omega,k}\lfloor_x$ is state-independent,
continuous in $x$, and actually it is determined by the local geometry of $(M,g)$ since
so is $\widetilde{\mathscr{G}}_k$ (by the Hadamard recursion relations).

Using the Leibniz rule, we can now rewrite the expression for $\omega(T^{ren}_{ab})$ as follows:
\begin{eqnarray}
 & & \omega(T^{ren}_{ab}(x)) = \omega(-:\phi \nabla_a \nabla_b\phi:{}_{\scSHP}(x) + \frac{1}{4}\nabla_a\nabla_b :\phi^2:{}_{\scSHP}(x)) \\
 & & \quad \quad + (\frac{1}{4} -\xi )\omega(\nabla_a\nabla_b :\phi^2:{}_{\scSHP}(x)) \nonumber \\
 & & \quad \quad + (4 \xi - 1) \left(\omega(-:\phi \nabla^c \nabla_c\phi:{}_{\scSHP}(x) + 
\frac{1}{4}\nabla^c\nabla_c :\phi^2:{}_{\scSHP}(x)) \right) g_{ab} \nonumber \\
 & & \quad \quad + \left(\left((1-4 \xi)(m^2+\xi R)-\frac{1}{2} \xi R \right)
g_{ab} - \xi G_{ab} \right) \omega(:\phi^2:{}_{\scSHP}(x)) \nonumber \\
 & & \quad \quad +\left( \left(4 \xi - \frac{1}{2}\right) F(x) - Q(x)\right) g_{ab}(x) 
+ C_{ab}(x)  \nonumber
\end{eqnarray}

By \eqref{BalDerDiffGeo} and once more the Leibniz rule, $\epsilon_{ab}$ can
also be expressed as
\begin{displaymath}
\epsilon_{ab} = \omega(-:\phi \nabla_a \nabla_b\phi:{}_{\scSHP}(x)
+ \frac{1}{4} \nabla_a \nabla_b :\phi^2:{}_{\scSHP}(x) )
\end{displaymath}
Thus, if $\omega$ is an $S_x^{(2)}$-thermal state of $\phi$, we obtain
\begin{eqnarray} \label{them}
& & \omega(T^{ren}_{ab}(x))  = \varepsilon_{ab}^\omega(x) + (4\xi - 1) g_{ab}(x)\varepsilon^\omega_c{}^c(x) \\
 &  & \quad \quad + (\frac{1}{4} -\xi)\nabla_a\nabla_b\vartheta^\omega(x) + (g_{ab}(x)\psi(x) -\xi R_{ab}(x)) \vartheta^\omega(x)
 \nonumber \\
  & & \quad \quad ((4\xi - 1/2)F(x)- Q(x))g_{ab}(x) + C_{ab}(x) \,, \nonumber
\end{eqnarray} 
where we use the abbreviation
\begin{equation}
\psi(x) = (1 - 4\xi)(m^2 + \xi R(x))\,.
\end{equation} 
With this expression, we are now in the position to derive bounds on  $v^av^b\omega(T^{ren}_{ab})$ for lightlike
or timelike vectors $v$. We will treat lower bounds first.
\begin{Thm} \label{ThmLB}
 Let $\phi$  be the quantized linear scalar field on $(M,g)$, with
parameters $m,\xi$ and $C_{ab}$, and let $\omega$ be a state of $\phi$ having two-point function
of Hadamard form. 
\begin{itemize}
\item[{\rm (a)}] 
 Suppose that $\xi = 1/4$, and let $v$ be a lightlike vector at $x \in M$, or a timelike vector
 at $x$ with $v_av^a =1$. If $\omega$ is in $\mathcal{L}_{\beta'}(x)$, $\beta' > 0$, then
 \begin{equation}
  v^av^b\omega(T_{ab}^{ren}(x)) \ge q(x,v;\beta')
 \end{equation} 
 where
\begin{eqnarray*}
q(x,v;\beta')  & = & - \frac{1}{4} \left| v^av^b R_{ab}(x) \right| \, \left| \frac{1}{{\beta'}^2}\chi_{0,m}(\beta') + c_{0,m}\right|
 \\
 & & + \left(\frac{F(x)}{2} - Q(x) - \frac{1}{4}c_{2,m}\right)v_av^a + v^av^bC_{ab} 
\end{eqnarray*}
\item[{\rm (b)}]
Let $\xi$ be arbitrary, let
$N \subset M$, and let $\gamma :[\tau_0,\tau_1] \to N$ be an affinely parametrized lightlike geodesic
defined on a finite interval, with tangent vector field
$v^a = \dot{\gamma}{}^a$. Suppose that $\eta$ is in $C_0^2((\tau_0,\tau_1))$ with $\eta \ge 0$. If $\omega \in \mathcal{L}_{\beta'}(N)$, there holds the bound
\begin{equation}
\int \eta(\tau)v^av^b\omega(T^{ren}_{ab}(\gamma(\tau)))\,d\tau
\ge q_0(\gamma,\eta;\beta')\,.
\end{equation}
Here, writing 
$$R_{[\gamma]} = \max_{\tau \in [\tau_0,\tau_1]} | \dot{\gamma}{}^a(\tau)\dot{\gamma}{}^b(\tau)R_{ab}(\gamma(\tau))|\,,$$
and defining $C_{[\gamma]}$ analogously, the bounding constant is given by
\begin{eqnarray*}
q_0(\gamma,\eta;\beta') & = & - \left[|\xi|R_{[\gamma]}\left|\frac{1}{{\beta'}^2}\chi_{0,m}(\beta') + c_{0,m}\right| 
                     + C_{[\gamma]} \right] ||\eta||_{L^1} \\
                      & & - |\frac{1}{4} - \xi|\cdot \left|\frac{1}{{\beta'}^2}\chi_{0,m}(\beta') + c_{0,m}\right|\cdot ||\eta''||_{L^1}
\end{eqnarray*}
\item[{\rm (c)}] Let $\xi$ be arbitrary, let $N \subset M$, and let $\gamma: [\tau_0,\tau_1] \to N$
be an affinely parametrized timelike geodesic with tangent vector field $v^a = \dot{\gamma}{}^a$, so that $v^av_a =1$.
Assume that  $\eta$ is in $C_0^2((\tau_0,\tau_1))$ with $\eta \ge 0$. If $\omega \in \mathcal{L}_{\beta'}(N)$, there holds the bound
\begin{equation}
\int \eta(\tau)v^av^b\omega(T^{ren}_{ab}(\gamma(\tau)))\,d\tau
\ge q_1(\gamma,\eta;\beta')\,,
\end{equation}
where, using the notation $\psi_{[\gamma]} = \max_{\tau \in [\tau_0,\tau_1]}|\psi(\gamma(\tau))|$, and
defining $F_{[\gamma]}$ and $Q_{[\gamma]}$ similarly, the bounding constant is given by
\begin{eqnarray*}
q_1(\gamma,\eta;\beta') & = &  - |\frac{1}{4} - \xi|\cdot \left|\frac{1}{{\beta'}^2}\chi_{0,m}(\beta') + c_{0,m}\right|\cdot ||\eta''||_{L^1}
\\
&  &  - (\psi_{[\gamma]} + |\xi| R_{[\gamma]})\cdot \left|\frac{1}{{\beta'}^2}\chi_{0,m}(\beta') + c_{0,m}\right|\cdot ||\eta||_{L^1} \\
&  & - \left( |4\xi - \mbox{$\frac{1}{2}$}| F_{[\gamma]} + |4 \xi - 1| |c_{2,m}| + Q_{[\gamma]} + C_{[\gamma]} + \frac{|c_{2,m}|}{4} \right) \cdot 
||\eta ||_{L^1}
\end{eqnarray*}
\end{itemize} 
\end{Thm}
{\it Proof}. The proof of the statement consists just of inserting the estimates of Lemma \ref{Lemma} and
discarding manifestly positive terms, in combination with estimate \eqref{estimate} for the average of the
second derivatives of $\vartheta^\omega$ along the geodesic. The term involving second derivatives of 
$\vartheta^\omega$ doesn't occur for $\xi = 1/4$, which makes it possible to give a pointwise lower bound
in this case. ${}$ \hfill $\Box$
\\[10pt]
The central assertion of Theorem \ref{ThmLB} is that the lower bound of the energy density averaged along
a causal geodesic depends only on the temperatures an LTE state attains on the geodesic, and is otherwise 
state-independent. The bound worsenes (shifts towards the left on the real axis) as
the temperature increases, i.e.\ with increasing $1/\beta'$.  It should also be noted that the bounds are local covariant.

For upper bounds on the averaged energy density of LTE states, an additional state-dependence shows up:
The bounds depend also on the tetrad $e$ appearing in the condition of $S_x^{(2)}$-thermality,
Def.\ \ref{Def-LTE-CST}. In this sense, the lower bounds on the averaged energy densities of LTE states are 
stronger than the upper bounds. This is similar to what holds for averages of energy densities for arbitrary Hadamard states
of the linear scalar field \cite{FewOst}.  

Let $x \in M$, and let $e = (e_0,\ldots,e_3)$ be an orthonormal tetrad at $x$ with $e_0$ timelike and future-pointing.
We define $\mathcal{L}_{\beta'}(x,e)$ as the set of all states $\omega$ in $\mathcal{L}_{\beta'}(x)$ where the
$S_x^{(2)}$-thermality conditions \eqref{beins} and \eqref{bzwei} hold with respect to the given tetrad.
Similarly, let $N$ be a subset of $M$, and let $N \owns x \mapsto e(x) = (e_0(x),\ldots,e_3(x))$ be a $C^0$
field of orthonormal tetrads over $N$, with  $e_0(x)$ timelike and future-pointing for all $x$. Then
we define $\mathcal{L}_{\beta'}(N,e)$ as the set of all states $\omega$ in $\mathcal{L}_{\beta'}(N)$ such
that, for each $x \in N$, $\omega$ satisfies the $S_x^{(2)}$-thermality conditions
\eqref{beins} and \eqref{bzwei} with respect to $e = e(x)$. With these conventions, we obtain the following 
upper bounds on (averaged) energy densities.
\begin{Thm} Let $\phi$ be the quantized linear scalar field on $(M,g)$, with 
parameters $m$, $\xi$ and $C_{ab}$, and let $\omega$ be a state of $\phi$ having
two-point function of Hadamard form.
\begin{itemize}
\item[(a)] Suppose that $\xi = 1/4$, let $v$ be a lightlike vector at $x \in M$, or a timelike vector at $x$
with $v_av^a = 1$, If $\omega$ is in $\mathcal{L}_{\beta'}(x,e)$, $\beta' > 0$, then
\begin{equation}
 p(v,x;\beta',e) \ge v^av^b\omega(T^{ren}_{ab}(x))
\end{equation}
where
\begin{displaymath}
\begin{split}p(v,x;\beta',e) = & \zeta(4)\frac{6(v^0)^2}{\pi^2{\beta'}^4} + \frac{1}{2}
\left| v^a v^b R_{ab} \right| \, \left| \frac{1}{{\beta'}^2} \chi_{0,m}(\beta')
+ c_{0,m} \right| \\ & + q(x,v;\beta')
\end{split}
\end{displaymath}
with $v^0 = v_a(e_0)^a$.
\item[(b)] Let $\xi$ be arbitrary, $N \subset M$, and let $\gamma:[\tau_0,\tau_1] \to N$
be an affinely parametrized lightlike geodesic defined on a finite interval, with tangent vector field $v^a = \dot{\gamma}{}^a$.
Suppose that $\eta$ is in $C^2_0((\tau_0,\tau_1))$ with $\eta \ge 0$. If $\omega$ is in $\mathcal{L}_{\beta'}(N,e)$,
then 
\begin{equation}
p_0(\gamma,\eta;\beta',e) \ge \int \eta(\tau)v^av^b\omega(T^{ren}_{ab}(\gamma(\tau))) d\tau 
\end{equation}
where
$$ p_0(\gamma,\eta;\beta',e) = \frac{6\zeta(4)}{\pi^2 (\beta')^2}(v^0_{[\gamma]})^2||\eta||_{L^1} + |q_0(\gamma,\beta';\eta)| $$ 
with $v^0_{[\gamma]} = \max_{\tau \in [\tau_0,\tau_1]}\dot{\gamma}{}_a(\tau) e_0^a(\gamma(\tau))$.
\end{itemize}
\end{Thm}

\section{Averaged Null Energy Condition (ANEC)}

In this section we derive the averaged null energy condition (ANEC) for
$S_N^{(2)}$-thermal states of the quantized linear scalar field $\phi$
on a globally hyperbolic spacetime $(M,g)$.

The ANEC on a state $\omega$ of $\phi$ demands that
\begin{equation} \label{eins1}
\liminf_{\tau_\pm \to \pm \infty}\,\int_{\tau_-}^{\tau_+} v^av^b \omega(T^{ren}_{ab}(\gamma(\tau))) \, d\tau \ge 0
\end{equation}
for all complete geodesics $\gamma$ in $M$ with affine parameter $\tau$ and
tangent $v^a = \dot{\gamma}{}^a$. If this condition holds, and if $(M,g)$
together with $\phi$ and $\omega$ are a solution to the semiclassical Einstein
equation in the form 
\begin{equation} \label{eins2}
G_{ab}(x) = 8\pi \omega(T^{ren}_{ab}(x))\,, \quad x \in M\,,
\end{equation}
then this implies that 
\begin{equation} \label{eins3}
\liminf_{\tau_\pm \to \pm \infty}\,\int_{\tau_-}^{\tau_+} v^av^b G_{ab}(\gamma(\tau)) \, d\tau \ge 0
\end{equation}
for all complete geodesics $\gamma$. (We address the issue for the semiclassical
Einstein equations with an additional contribution by a classical stress-energy tensor below.)
It has been shown that this weaker form of the usual pointwise null energy condition,
which demands that $\ell^a \ell^b G_{ab}(x) \ge 0$ for all lightlike vectors $\ell^a$
at each $x \in M$, is still sufficient to reach the same conclusions with respect to 
singularity theorems as obtained from the pointwise null energy condition, i.e.\ that congruences
of geodesics will focus with expansion diverging to $-\infty$ at finite affine geodesic parameter \cite{HawkEllis}. The validity of \eqref{eins1} is therefore of importance
for the properties of the spacetime structure of solutions to the semiclassical
Einstein equations.

It has been argued
in \cite{YurWal} that condition \eqref{eins1} may be replaced by the following condition:
\begin{equation} \label{zwei1}
\liminf_{\lambda \to 0} \int_{-\infty}^{\infty} \eta_\lambda(\tau)v^av^b\omega(T^{ren}_{ab}(\gamma(\tau)))
\,d\tau \ge 0
\end{equation}
for any $\eta \in C_0^2(\mathbb{R})$, $\eta \ge 0$, with $\eta(0) > 0$ and
$\eta_{\lambda}(\tau) = \eta(\lambda \tau)$ for $\lambda > 0$. More precisely,
in \cite{YurWal} it has been shown that \eqref{zwei1} and \eqref{eins2} imply that
the expansion of a congruence of lightlike geodesics around $\gamma$ becomes
singular along $\gamma$ (in the sense of diverging to $-\infty$ at a 
finite value of the affine parameter) unless it 
vanishes identically on $\gamma$. (In \cite{YurWal}  this argument
is given for half-line geodesics, but it carries over to the case at hand as will be shown
in our Appendix A.) 

Now let $\omega \in S^{(2)}_N$, and let $\gamma$ be a complete lightlike geodesic in $N \subset M$
with affine parameter $\tau$ and tangent $v^a = \dot{\gamma}{}^a$. Then, from \eqref{them},
\begin{equation} \label{zwei2}
v^av^b\omega(T^{ren}_{ab}) = v^av^b\varepsilon_{ab}^\omega + \left(\frac{1}{4} - \xi\right)v^av^b\nabla_a\nabla_b
\vartheta^\omega - \xi v^av^bG_{ab} \vartheta^\omega + v^av^b C_{ab}
\end{equation}
holds along $\gamma$. Therefore, positivity properties of the (integrated) energy
density $v^av^b\omega(T^{ren}_{ab})$ depend also on the behaviour of $G_{ab}$ and
$C_{ab}$. The sign of the term involving $G_{ab}$ is not known. To circumvent this
difficulty, we assume that the underlying spacetime $(M,g)$ together with
$\phi$ and $\omega$ are solutions to the semiclassical Einstein equations \eqref{eins2},
since it is this situation in which the ANEC is applied to deduce \eqref{eins3} and the
ensueing statements about focussing of lightlike geodesics. Supposing that $(M,g)$ together with
$\phi$ and $\omega$ are solutions to the semiclassical Einstein equations, and also that
$\omega$ is an $S^{(2)}_N$-thermal state, we obtain upon combination of
\eqref{eins2} and \eqref{zwei2} the equation
\begin{equation} \label{drei1}
v^av^b[G_{ab}(1 + 8\pi\xi \vartheta^\omega) - 8\pi C_{ab}] = 
 8\pi v^av^b\left(\varepsilon_{ab}^\omega + \left(\frac{1}{4} - \xi\right) \nabla_a\nabla_b \vartheta^\omega\right)
\end{equation}
on $N$.
In order to draw further conclusions, one must specify $C_{ab}$. We recall that $C_{ab}$ is a
datum of the linear quantum field $\phi$, a priori only restricted by the requirement that
$T^{ren}_{ab}$ be a local covariant quantum field and divergence-free, thus $C_{ab}$ should be locally constructed
from the spacetime metric. Following Wald \cite{WaldQFT}, one can make the assumption that 
$C_{ab}$ have canonical dimension, which leads to the form
\begin{equation} \label{vier1}
 C_{ab} = A  g_{ab} + B G_{ab} + \Gamma \frac{\delta}{\delta g^{ab}} S_1(g)+ D \frac{\delta}{\delta g^{ab}}S_2(g) 
\end{equation}
where $S_1(g) = \int_M R^2 d{\rm vol}_g$, $S_2(g) =
\int_M R_{ab}R^{ab}d{\rm vol}_g$, and $\delta/\delta g^{ab}$
means functional differentiation with respect to the metric,
with constants $A$, $B$, $\Gamma$, $D$ as remaining renormalization ambiguity for the
quantum field $\phi$ (see \cite{WaldQFT} for additional discussion). For the rest of our discussion,
we will simplify matters by assuming $\Gamma,D =0$.

Making this assumption, so that \eqref{vier1} holds, and observing that hence,
$v^av^bC_{ab} = B v^av^bG_{ab}$ for all lightlike vectors $v^a$, \eqref{drei1}
assumes on $N$ the form
\begin{equation} \label{vier2}
v^av^bG_{ab}(1 + 8\pi(\xi \vartheta^\omega -B)) =
 8\pi v^av^b\left(\varepsilon_{ab}^\omega + \left(\frac{1}{4} - \xi\right) \nabla_a\nabla_b \vartheta^\omega\right)\,.
\end{equation}
The constant $B$ is still free, and one may now try to choose $B$ in such a way that 
\eqref{vier2} entails the ANEC for all lightlike geodesics in $N \subset M$ and an as large as
possible class of $S^{(2)}_N$-thermal states $\omega$. We will show that this is possible with different conditions on $B$ for
the cases $\xi = 1/4,0<\xi<1/4,\xi=0$.

\begin{Thm} \label{ThmANEC}
Let $(M,g)$ be a globally hyperbolic spacetime, let $\phi$ be the quantized linear
scalar field on $(M,g)$, with parameters $m,\xi,C_{ab}$, where
$C_{ab} = A g_{ab} + B R_{ab}$, with real constants $A,B$.

Suppose further that $\omega$ is a quasifree Hadamard state for $\phi$, that $\omega \in S^{(2)}_N$
for $N \subset M$,
and that $(M,g)$ together with $\phi$ and $\omega$ provides a solution to the semiclassical
Einstein equation \eqref{eins2}. 

Let $\gamma$ be a complete lightlike geodesic in $N$ with affine parameter $\tau$ and
tangent $v^a = \dot{\gamma}{}^a$, and let $\eta \in C_0^2(\mathbb{R})$, $\eta \ge 0$. Then
\begin{equation} \label{funf1}
\lim_{\lambda \to 0} \, \int_{-\infty}^{\infty} \eta(\lambda \tau) \omega(T^{ren}_{ab}(\gamma(\tau)))\,d\tau
\ge 0
\end{equation}
holds if any of the following groups of conditions is assumed:
\\[6pt]
{\rm 1.)} \quad  $\xi = 1/4$, $B < 1 + 2\pi c_{0,m}$. In this case one even has
$$ v^av^b\omega(T_{ab}^{ren}(x)) \ge 0 $$
pointwise for all $x \in M$ and all lightlike vectors $v^a$ at $x$.
\\[6pt]
{\rm 2.)} \quad $0 < \xi < 1/4$, $B \leq \xi c_{0,m} + 1/(8\pi)$, 
\begin{eqnarray} \label{sechs1}
& &\lambda \ln(\vartheta^\omega(\gamma(\tau/\lambda))) \to 0 \ \ \text{as} \ \ \lambda \to 0 \ \ \text{for almost all}\ \ \tau\,, \\ \label{sechs2}
& & \int_s^r \lambda |\ln(\vartheta^\omega(\gamma(\tau/\lambda)))|\,d\tau < k < \infty \ \
\text{for small}\ \lambda \ \ \text{and all} \ s < r \in \mathbb{R}\,.
\end{eqnarray}
{\rm 3.)} \quad $\xi = 0$, $B < 1/8\pi$,
\begin{eqnarray} \label{sechs3}
& &\lambda \vartheta^\omega(\gamma(\tau/\lambda)) \to 0 \ \ \text{as} \ \ \lambda \to 0 \ \ \text{for almost all}\ \ \tau\,, \\ \label{sechs4}
& & \int_s^r \lambda \vartheta^\omega(\gamma(\tau/\lambda))d\tau < K < \infty \ \
\text{for small}\ \lambda\ \ \text{and all} \ s < r \in \mathbb{R}\,.
\end{eqnarray}
\end{Thm}
{\bf Remark} \ \ (a) \ \ If, instead of \eqref{eins2}, the semiclassical Einstein equations are
assumed to hold in the form
$$ G_{ab}(x) = 8\pi (T_{ab}^{class}(x) + \omega(T_{ab}^{ren}(x))) $$
with a stress-energy tensor $T^{class}_{ab}$ for classical, macroscopic matter
distribution, and if it is assumed that this stress-energy tensor fulfills the
pointwise null energy condition $\ell^a\ell^b T_{ab}^{class}(x) \ge 0$ for all lightlike
vectors $\ell^a$ at each point $x \in M$, then the statements of the theorem remain valid with
$T_{ab}^{class} + \omega(T_{ab}^{ren})$ in place of $\omega(T^{ren}_{ab})$.
\\[6pt]
(b) \ \ Conditions \eqref{sechs1} and \eqref{sechs2} say, roughly speaking, that 
$\vartheta^\omega(\gamma(\tau))$ shouldn't grow faster than ${\rm e}^{|\tau|^{(1 - \epsilon)}}$ for
$|\tau| \to \infty$, while \eqref{sechs3} and \eqref{sechs4} say that $\vartheta^\omega(\gamma(\tau))$
shouldn't grow faster than $|\tau|^{1-\epsilon}$ as $|\tau| \to \infty$. Now since
$\vartheta^\omega(\gamma(\tau)) = (\beta(\gamma(\tau))^{-2}\chi_{0,m}(\beta(\gamma(\tau))) + c_{0,m}$
and since
$$ \chi_{0,m}(\beta) \to \frac{1}{2 \pi^2} \int_0^\infty \frac{\rho}{{\rm e}^\rho}\,d\rho 
\quad \text{for} \ \ \beta \to 0\,,$$
this means that the growth of the temperature $1/\beta(\gamma(\tau))$ at $\gamma(\tau)$
appearing in Def.\ \ref{Def-LTE-CST} of $S^{(2)}_{\gamma(\tau)}$-thermality $\omega$ should
not exceed ${\rm e}^{|\tau|^{(1 - \epsilon)/2}}$ and $|\tau|^{(1 - \epsilon)/2}$ as 
$|\tau| \to \infty$, respectively.
\\[10pt]
{\it Proof of Thm.\ \ref{ThmANEC}.} 1.)\ \  If $\xi = 1/4$, then \eqref{vier2} assumes the form 
\begin{equation} \label{sechs5}
v^av^bG_{ab}(1 + 8\pi( \vartheta^\omega/4  - B)) = 8\pi v^av^b\varepsilon^\omega_{ab}\,.
\end{equation}
If $B < 1 + 2\pi c_{0,m}$, then the factor $1 + 8\pi(\vartheta^\omega/4 - B)$ is strictly
positive, as is the right hand side of \eqref{sechs5}. This equality holds pointwise at all
$x \in M$ and for all lightlike vectors $v^a$, thus proving, in combination with the assumed property
\eqref{eins2}, the statement of the theorem.
\\[6pt]
2.)\ \  For $0< \xi < 1/4, B=\xi c_{0,m}+1/(8 \pi) - \xi c$, where $c\geq 0$,
\eqref{vier2} takes the form
\begin{equation} \label{sechs6}
v^av^b G_{ab} (8\pi \xi (\vartheta^\omega -c_{0,m} + c)) = 8\pi v^a v^b \varepsilon_{ab}^\omega + 8 \pi (1/4-\xi)
v^a v^b \nabla_a\nabla_b \vartheta^\omega\,.
\end{equation}
Observing that $v^a v^b\nabla_a \nabla_b c_{0,m} = 0$, the last equation is 
turned into
\begin{equation} \label{sieben1}
v^a v^b G_{ab} = \frac{v^a v^b \varepsilon^\omega_{ab}}{\xi (\vartheta^\omega - c_{0,m} + c)} + 
 \frac{(1/4 - \xi) v^a v^b \nabla_a \nabla_b(\vartheta^\omega - c_{0,m})}
{\xi (\vartheta^\omega - c_{0,m} + c)}
\end{equation}
where it was used that $\vartheta^\omega - c_{0,m} + c> 0$. The first term on 
the right hand side of \eqref{sieben1} is positive. Upon integration against 
a non-negative $C_0^2$ weighting function $\eta$ along the geodesic $\gamma$ 
we obtain, using the abbreviation
$$ u(\tau) = \vartheta^\omega(\gamma(\tau)) - c_{0,m}\,,$$
the inequality
$$  \int \eta(\tau) (v^av^b G_{ab})(\gamma(\tau)) \, d\tau 
\ge \frac{1/4-\xi}{\xi} \int \eta(\tau) \frac{u''(\tau)}{u(\tau)+c}\,d\tau\,.$$
By partial integration,
$$ \int \eta(\tau) \frac{u''(\tau)}{u(\tau)+c}\,d\tau = \int\eta(\tau) \left(\frac{u'(\tau)}{u(\tau)+c} \right)^2
 \,d\tau + \int \ln(u(\tau)+c) \eta''(\tau)\,d\tau\,.$$
Thus, since the first integral on the right hand side is non-negative,
$(1/4-\xi)/\xi>0$ for the $\xi$ considered and using the monotonicity of the 
logarithm together with $c \geq 0$,
$$ \int \eta(\lambda \tau) (v^av^b G_{ab})(\gamma(\tau)) \, d\tau
\ge \frac{1/4-\xi}{\xi} \int \lambda \ln(u(\tau/\lambda)) \eta''(\tau)\,d\tau \,,$$
and owing to assumptions \eqref{sechs1} and \eqref{sechs2}, the expression on the right hand side
converges to $0$ as $\lambda \to 0$.
Equation \eqref{funf1} is then again implied by the assumed property \eqref{eins2}.
\\[6pt]
3.)\ \ If $\xi = 0$, equation \eqref{vier2} turns into
\begin{equation} \label{acht1}
v^a v^b G_{ab}(1 - 8\pi B) = 8\pi v^av^b \varepsilon^\omega_{ab}
 + \frac{1}{4} v^av^b \nabla_a\nabla_b \vartheta^\omega\,,
\end{equation}
and by the condition on $B$, the factor $1 - 8\pi B$ is strictly positive. Observing again positivity
of $8\pi v^av^b \varepsilon_{ab}^\omega$, upon integration against a non-negative $C_0^2$
weighting function $\eta$ along $\gamma$ one obtains
$$ \int \eta(\lambda \tau) v^av^b G_{ab}(\gamma(\tau)) \, d\tau
\ge \frac{1}{4(1 -8\pi B)}\int \lambda u(\tau/\lambda) \eta''(\tau)\,d\tau $$
and the right hand side converges to $0$ as $\lambda \to 0$ by assumptions
\eqref{sechs3} and \eqref{sechs4}. Again \eqref{funf1} is deduced from the assumed validity of
\eqref{eins2}. ${}$ \hfill $\Box$

\section{Generalized Local Thermal Equilibrium States}

The notion of LTE states in \cite{BuOjiRoo}, and the related definition of
$S^{(2)}_x$-thermal states, is actually more general than the definition given in Sec.\ 2.
In \cite{BuOjiRoo} the possibility was considered that an LTE state $\omega$ coincides on
$S_x$-observables not necessarily with a thermal equilibrium state at sharp temperature
in a certain Lorentz frame, but with a mixture of such states.

In our setting, where we work with the linear scalar field, this corresponds to
a modification of Def.\ 2.3 as follows.
As a consequence of eqn.\eqref{equili}, $\omM^{\beta e}$,
the quasifree thermal equilibrium state with respect to the Minkowski tetrad
$e = (e_0,e_1,e_2,e_3)$ at inverse temperature $\beta$, depends only on
$\bB = \beta e_0$. This quantity completely parametrizes $\omM^{\beta e}$, so we write
$\omM^{\bB}$ in place of $\omM^{\beta e}$. The vectors $\bB$ take values in $V^+$, the set of
future-directed timelike vectors in Minkowski spacetime. 

Let $(M,g)$ be a globally hyperbolic spacetime, let $V_x^+ \subset T_xM$ be the set of
future-directed timelike vectors at $x \in M$, and let $\rho_x$ be a Borel measure on
$V_x^+$ supported on a compact subset $B_x \subset V_x^+$, with $\int_{B_x} d\rho_x(\bB) =1$.
Then we say that a Hadamard state $\omega$ of the linear scalar field $\phi$ on $(M,g)$ is
a {\it generalized} $S^{(2)}_x$-{\it thermal state} if
\begin{eqnarray}
\omega(:\phi^2:{}_{\scSHP} (x)) 
                          & = & \int_{B_x}\omM^{\bB}(:\phM^2:(\xM))
d\rho_x(\bB) + c_{0,m}\,, \nonumber \\
v^aw^b\omega(\balder_{ab}:\phi^2:{}_{\scSHP} (x))
& = & v^\mu w^\nu \int_{B_x} \omM^{\bB}(\balder_{\mu\nu}:\phM^2:(\xM))
d\rho(\bB) + c_{2,m} v^\mu w^\nu \eta_{\mu\nu} \nonumber
\end{eqnarray}
holds for all (spacelike) vectors $v,w \in T_xM$  for
some $\xM \in \MM$.  
Making further the assumption that $F \mapsto \int_M \int_{B_x} F(x,\bB) d\rho_x({\bB}) d{\rm vol}(x)$,
$F \in C_0(TM,\mathbb{C})$ is a distribution (on the manifold $TM$), such that
$x \mapsto \int_{B_x} F(x,\bB) d\rho_x(\bB)$ is $C^2$, one can define
generalized $S_N^{(2)}$-thermal states in analogy to the  definition of
$S^{(2)}_N$-thermal states in Sec.\ 2. 

With these conventions and assumptions, the results of Thms.\ 3.1, 3.2 and 4.1 extend to
generalized $S_N^{(2)}$-thermal states, under identical assumptions, except that
 the bounds have to be corrected for the $\rho_x$-integrations. It should be obvious
how this is to be done.

\section{Discussion and Outlook}

We have generalized the concept of local thermal equilibrium states of \cite{BuOjiRoo}, or rather,
the concept of $S^{(2)}_x$-thermal states, to the quantized linear scalar field models
on generic globally hyperbolic spacetimes, and have shown that one can derive certain
quantum energy inequalities for such states. The lower bounds appearing in the quantum
energy inequalities of local thermal equilibrium states depend only on the local temperature
of the states, i.e.\ thermal function $\vartheta^\omega$, corresponding to the expectation
value of the Wick-square in local thermal equilibrium states. The upper bounds, instead, 
depend also on the local frames with respect to which $S^{(2)}_x$-thermality is defined.
In this sense, the lower bounds are stronger (have less dependence on the states) than the
upper quantum energy inequality bounds. This is a feature also found for quantum energy inequalities
of general Hadamard states of the linear scalar fields, and has led to the proposal to 
consider `relative quantum energy inequalities' as a more general variant of quantum energy inequalities which has the potential to be valid also in interacting quantum field theories \cite{FewSmith}.
Moreover, the quantum energy inequalities for thermal equilibrium states are local covariant.

A major purpose of quantum energy inequalities, especially in local covariant form, is to provide information about the structure
of spacetime geometries appearing as solutions to the semiclassical Einstein equations. Quite
generally, they serve as stability conditions on quantum matter, and ensure that correspondingly
the (semiclassical) gravitational interaction is attractive, at least when averaged over
sufficiently extended spacetime regions. The averaged null energy condition which we proved for
certain values of the curvature coupling $\xi$ and certain values of the renormalization constants
is of a similar nature. One may also take the requirement that the ANEC should be fulfilled
for suitable thermal equilibrium states
as a constraining condition on the largely free choice of renormalization constants for the
stress-energy tensor. Certainly a demand in this spirit leads to further relations between
the renormalization constants, the parameters fixing the field model, and possibly geometrical
quantities, and for this reason it is attractive to further study quantum energy inequalities and
ANEC in the context of solutions to the semiclassical Einstein equations.

One important issue we haven't addressed at all so far is the existence of local thermal equilibrium
states, or at least $S^{(2)}_N$ states for subsets $N$ in
spacetime. We have simply assumed that there are such local thermal equilibrium states
to which our results apply. The question if there are local thermal equilibrium states in generic spacetimes is an interesting and difficult problem, for which we can't offer, as yet,
any route to its solution. 
However, the existence of LTE states for the massless and the massive Klein-Gordon fields
on (parts of) Minkowski spacetime has been established, with an interesting relation 
to situations resembling a big bang scenario \cite{Bu,Hueb}.

The question if local thermal equilibrium states exist is a first step towards the question
how generic they are. One is inclined to think that within certain time- and energy scales,
local thermal equilibrium states should be the archetypical physical states in the sense that,
if one is asked to randomly pick a state in the physical state space (of a quantum field
theory), then the result would be a local thermal equilibrium state with overwhelming
likelihood. At least this is expected for interacting quantum field theories since interaction
tends to equilibrate subsystems (or degrees of freedom) of a large system. If this turned out
to be true, and if the temperature distribution of such states turned out to
allow ANEC results similar to those of Thm.\ 4.1, then one would be led to conclude
that (under general additional assumptions) the occurrence of singularities in solutions
to the semiclassical Einstein equations is a generic feature. It would be of utmost interest
to investigate this circle of questions further particularly in scenarios of early cosmology.  
\\ \\
{\bf Appendix}
\\[10pt]
{\bf A} We will present a result on real-valued solutions $\theta(t)$ of the
differential equation 
\begin{equation} \label{diffeq}
\theta'(t) + \mu \theta(t)^2 = - f(t) \,, \quad t \in \mathbb{R}\,,
\end{equation}
where $\mu > 0$ and $f \in C^1(\mathbb{R},\mathbb{R})$, with initial condition
\begin{equation} \label{AWP}
\theta(0) = \theta_0\,.
\end{equation}
 It follows from the Picard-Lindel\"of Theorem that
there is an open interval $(a,b)$ containing $0$,
 which may be finite, semi-finite or infinite
(i.e.\ coinciding with $\mathbb{R}$), such that this interval is the domain of
the unique, inextensible $C^1$ solution $\theta$ of \eqref{diffeq} satisfying the
initial condition. In this case, we call $\theta$ the maximal solution of 
\eqref{diffeq} defined by the initial condition, and refer to $(a,b)$ as the
maximal domain.

The following statement is a variation on a similar result in \cite{YurWal}, 
and it uses a very similar argument, the main difference being that the 
assumption \eqref{lowbound} here is slightly different from that in 
\cite{YurWal}, where the integral is taken over a semi-axis. Note also that 
our parameter $\lambda$ corresponds to $1/\lambda$ in the notation of 
\cite{YurWal}.
\\[10pt]
{\bf Theorem A.1} \ \
\it 
Suppose that $f \in C^1(\mathbb{R},\mathbb{R})$ has the
property 
\begin{equation} \label{lowbound}
 \limsup_{\lambda \to 0}\,\int_{-\infty}^{\infty} f(t)\eta(\lambda t)\,dt  \ge 0
\end{equation}
for the function $\eta(t) = (1- t^2)^4$ for $|t| < 1$, $\eta(t) = 0$ for $|t| \ge 1$.

Then either the maximal domain of $\theta$ coincides with all of the real axis and
$\theta(t) = 0$ for all $t \in \mathbb{R}$, or the maximal domain $(a,b)$ of $\theta$ 
is a finite or semi-finite interval. In this case, $\theta(t) \to 
\mp \infty$ for $t$ approaching the finite boundary at the right/left side
of the maximal domain (in the finite case this holds with the respective
sign for both boundaries). In particular, this is the case if 
$\theta(t_0) \ne 0$ for some $t_0$ in the maximal domain of $\theta$. 
\\[6pt]
{\it Proof.} 
\rm
Consider the auxiliary differential equation
\begin{equation}
u''(t) + \frac{f(t)}{\mu} u (t) = 0
\label{auxeqn}
\end{equation}
For the initial values $u(0)=1$, $u'(0) = \theta_0$ and the given $f$ this 
linear differential equation has by the Picard-Lindel\"of Theorem a unique, 
global solution  $u \in C^2(\R, \R)$. 
Furthermore, this solution is nonzero in some neighbourhood of $0$. For 
points from this neighbourhood, one can then rewrite \eqref{auxeqn} as
\begin{displaymath}
\frac{\mathrm{d}}{\mathrm{d}t} \left( \frac{u'(t)}{u(t)} \right)
+ \left( \frac{u'(t)}{u(t)} \right)^2 = - \frac{f(t)}{u(t)}
\end{displaymath}
which implies that $\hat{\theta}(t) \equiv \frac{u'(\mu t)}{u(\mu t)}$
fulfills equation \eqref{diffeq}. Furthermore, $\hat{\theta}$ also satisfies
the initial condition \eqref{AWP} and by the uniqueness part in the
Picard Lindel\"of Theorem it therefore agrees with $\theta$. This however
implies that the only way in which $\theta$ can fail to be $C^1$ at a
boundary point $c=a$ or $c=b$ of a semi-finite interval is a zero of 
$u$ at $\mu c$. At this zero $u'$ has to differ from zero, 
otherwise $u$ as a $C^2$-solution to \eqref{auxeqn} with initial conditions
$u(\mu c)=0, u'(\mu c)=\lim_{x \to \mu c} u'(x) = 0$ would be identically zero in 
contradiction to the initial values for $u$ at $0$. By continuity, $u'$ is 
therefore  nonzero in a neighbourhood of $\mu c$, and by \eqref{diffeq}, $ \theta(t) = \frac{u'(\mu t)}{u(\mu t)}$ 
approaches the value $- \infty$ for $t \to c, t <c$ (right boundary point)
or the value $+ \infty$ for $t  \to c,  t > c$ (left boundary point). For proving that 
$\theta$ diverges at the boundary (boundaries) of a semi-finite interval it 
is therefore sufficient to show that $\theta$ cannot be continued as a $C^1$ 
function beyond this boundary. 

With the definition of $\eta$ as above, and provided that the maximal
domain of $\theta$ coincides with all of $\mathbb{R}$, one has for $0 < \lambda < 1$,
\begin{eqnarray*}
 \int_{-\infty}^{\infty} \theta'(t) \eta(\lambda t)\,dt
. &=& \int_{-\infty}^{\infty} \theta(t) \lambda \eta'(\lambda t)\,dt \\
  &=& -8\lambda\int_{-1/\lambda}^{1/\lambda} \theta(t) (\lambda t)(1 - (\lambda t)^2)^3\,dt \\
  &\ge &  -8\lambda \int_{-1/\lambda}^{1/\lambda}|\theta(t)|\,(1 -(\lambda t)^2)^2\,dt
\end{eqnarray*}
owing to the fact that both $|\lambda t|$ and $|(1 - (\lambda t)^2)|$ are bounded by $1$ on
the domain of integration. Combining this with \eqref{diffeq} and \eqref{lowbound}
leads to
\begin{equation} \label{interm}
\limsup_{\lambda \to 0} \, -8\lambda \int_{-1/\lambda}^{1/\lambda} |\theta(t)|(1 -(\lambda t)^2)^2 \,dt + \mu \int_{-1/\lambda}^{1/\lambda} \theta(t)^2(1 - (\lambda t)^2)^4\,dt \le 0\,.
\end{equation}
Using also the Cauchy-Schwarz inequality
$$ \int_{-1/\lambda}^{1/\lambda} |\theta(t)|(1 -(\lambda t)^2)^2\,dt
 \le \left(\int_{-/\lambda}^{1/\lambda} \theta(t)^2(1- (\lambda t)^2)^4\,dt\right)^{1/2}
     \left( \int_{-1/\lambda}^{1/\lambda} 1\,dt \right)^{1/2}\,,$$
the estimate \eqref{interm} can be replaced by
\begin{equation}
\limsup_{\lambda \to 0}\,
-\frac{8}{\mu}\sqrt{2 \lambda} \left(\int_{-1/\lambda}^{1/\lambda} \theta(t)^2(1- (\lambda t)^2)^4\,dt\right)^{1/2} + \int_{-1/\lambda}^{1/\lambda} \theta(t)^2(1- (\lambda t)^2)^4\,dt
   \le 0\,,
\end{equation}
which shows that $\int_{-\infty}^{\infty} \theta(t)^2\,dt = 0$ upon using Levi's theorem.
Since $\theta$ is $C^1$, this implies that $\theta(t) = 0$ for all $t$. 

We have therefore shown that the assumption of $\theta$ being $C^1$ on all of $\mathbb{R}$
implies $\theta(t)=0$ for all $t \in \R$; if on the other hand $\theta$
is $C^1$ only on a maximal finite or semi-finite interval, then by the statement in the 
first paragraph of the proof, it will diverge at the finite boundaries
of this interval in the indicated way.
\\[20pt]
{\bf B} Here we will calculate the constants $c_{0,m}, c_{2,m}$ that
arise when defining the Wick-square and the second balanced derivative
on Minkowski spacetime using the covariant point-split renormalization. 
A similar calculation can also be found in the Appendix B of \cite{Moretti},
the different conventions adapted here however lead to small changes
in some of the formulas appearing. \\
The Hadamard recursion-relations satisfied by the functions $U_j$ in 
\eqref{HadVDefn} read with our sign-conventions:
\begin{align*}
- 2  (\nabla^{\kappa} \sigma) \nabla_{\kappa} U_0 - (4 + \nabla^{\kappa}
\nabla_{\kappa} \sigma) U_0 & = 
(\nabla^{\kappa} \nabla_{\kappa} + m^2 + \xi R) U \\
- 2 (\nabla^{\kappa} \sigma) \nabla_{\kappa} U_{j+1} + 
( 4j - \nabla^{\kappa} \nabla_{\kappa} \sigma) U_{j+1}
& = \frac{(\nabla^{\kappa} \nabla_{\kappa} + m^2 + \xi R) U_j}{j+1}
\end{align*}
For Minkowski spacetime, $U$ is identically one, $\nabla^{\kappa}
\nabla_{\kappa} \sigma = -8$ and the unique solutions of the resulting
recursion relations
\begin{align*}
4 (x-x')^{\kappa} \nabla_{\kappa} U_0 + 4 U_0 & = m^2 \\
4 (x-x')^{\kappa} \nabla_{\kappa} U_{j+1} + 4 (2+j) U_{j+1} & = \frac{(m^2 + \nabla^{\kappa} \nabla_{\kappa}) U_j}{j+1}
\end{align*}
that remain bounded for $x \to x'$ are easily calculated (e.g. using
the method of characteristics) as
\begin{displaymath}
U_j = \frac{1}{j! (j+1)!} \left( \frac{m^2}{4} \right)^{j+1}
\end{displaymath}
For non-lightlike $x-x'$ where $G_{k,\epsilon}$ is a regular distribution
(the corresponding function being obtained as the pointwise limit 
$\epsilon \to 0$) we have with the abbreviation 
$(x-x')^2 := \eta_{ab} (x-x')^a (x-x')^b$:
\begin{displaymath}
G_{1,0}(x,x') = \frac{1}{4 \pi^2} \left( \frac{1}{-(x-x')^2}
+ \frac{m^2}{4} \ln \left( -(x-x')^2 \right) 
\left[1+\frac{-m^2(x-x')^2}{8} \right] \right)
\end{displaymath}
(it will be seen in the course of the calculation, that $G_{1,0}$ is
actually sufficient to calculate the second balanced derivative, one does
not need $G_{2,0}$). The two-point function $W_2^{\omega^{\rm vac}}$ of the 
Minkowski vacuum state $\omega^{\rm vac}$ for spacelike $(x-x')$ is given by 
\cite{Bogo}
\begin{displaymath}
W_2^{\omega^{\rm vac}}(x,x') = \frac{m}{4 \pi^2} 
\frac{K_1 \left(m \sqrt{-(x-x')^2}\right)}{\sqrt{-(x-x')^2}}
\end{displaymath}
and using the asymptotic expansion of the modified Bessel function $K_1$ for
small arguments, the terms up to the order $(x-x')^2$ of the two-point
function are given by
\begin{displaymath}
\begin{split}
W_2^{\omega^{\rm vac}} (x,x') = \frac{1}{4 \pi^2} \bigg(&
\frac{1}{-(x-x')^2}+ \frac{m^2}{4} \ln \left( \frac{-m^2 (x-x')^2}{4} \right)
\left[ 1 - \frac{m^2}{8} (x-x')^2 \right] \\ & +
\frac{m^2}{4} \left[
(2 \gamma -1) + (2 \gamma - 5/2) \frac{- m^2 (x-x')^2}{8} \right] \bigg)
\end{split}
\end{displaymath}
(here and in the following, $x-x'$ is now assumed to be spacelike). The 
difference $W_2^{\omega^{\rm vac}}(x+\zeta,x-\zeta) - G_{1,0}(x+\zeta,x-\zeta)$
to the order required for the calculation of
$\omega^{\infty}(:\phi^2:{}_{\scSHP}(x))$
and $\omega^{\infty}(\balder_{\mu \nu}:\phi^2:{}_{\scSHP}(x))$
is then
\begin{displaymath}
\begin{split}
W_2^{\omega^{\rm vac}}(x+\zeta,x-\zeta) - G_{1,0}(x+\zeta,x-\zeta)
= & \frac{m^2}{(4 \pi)^2} \bigg[ \ln \left( \frac{m^2}{4} \right) 
+ 2 \gamma - 1 \\ & + \left( \ln \left( \frac{m^2}{4} \right) 
+ 2 \gamma  - 5/2 \right) \frac{-m^2 \zeta^2}{2} \bigg]
\end{split}
\end{displaymath}
With this expression one calculates
\begin{align*}
\omega^{\infty}(:\phi^2:{}_{\scSHP}(x)) & = \frac{m^2}{(4 \pi)^2}
\left[ \ln \left( \frac{e^{2 \gamma} m^2}{4} \right) -1 \right] =: c_{0,m}\\
\omega^{\infty}(\balder_{\mu \nu}:\phi^2:{}_{\scSHP}(x)) & = 
- \frac{m^4}{(4 \pi)^2} \left[ \ln \left( \frac{e^{2 \gamma} m^2}{4} \right) 
- \frac{5}{2} \right] \eta_{\mu \nu} =: c_{2,m} \eta_{\mu \nu}
\end{align*}
and from this one reads of the equations \eqref{Wickdiff1} and 
\eqref{Wickdiff2}. 
\\[24pt]
{\bf Acknowledgements} \quad  The authors would like to thank D.\ Buchholz for discussions on 
local thermal equilibrium states. J.S.\ gratefully acknowledges financial support
by the International Max Planck Research School (IMPRS).


\begin{thebibliography}{22}

\newcommand{\enquote}[1]{``#1''}
\ifdefined\href
\providecommand{\eprint}[1]{\href{http://www.arxiv.org/abs/#1}{\texttt{#1}}}
\else
\providecommand{\eprint}[1]{\texttt{#1}}
\fi

\bibitem{Alcub}
Alcubierre M., \enquote{{The Warp Drive: Hyper-Fast Travel within General
  Relativity}}, Class. Quant. Grav. \textbf{11} (1994) L73,
  \eprint{gr-qc/0009013}

\bibitem{Baeretal}
B{\"a}r C., Ginoux N., Pf{\"a}ffle F., \emph{Wave Equations on {L}orentzian
  Manifolds and Quantization}, ESI Lectures in Mathematics and Physics,
  European Mathematical Society (EMS), Z\"urich (2007)

\bibitem{Bogo}
Bogolubov N.N., Logunov A.A., Oksak A.I., Todorov I.T., \emph{General
  Principles of Quantum Field Theory}, vol.~10 of \emph{Mathematical Physics
  and Applied Mathematics}, Kluwer Academic Publishers Group, Dordrecht (1990)

\bibitem{CasimEng2}
Bordag M., Mohideen U., Mostepanenko V.M., \enquote{New Developments in the
  Casimir Effect}, Phys. Rept. \textbf{353} (2001) 1

\bibitem{GFBorde}
Borde A., \enquote{{Geodesic Focusing, Energy Conditions and Singularities}},
  Class. Quant. Grav. \textbf{4} (1987) 343

\bibitem{BraRo}
Bratteli O., Robinson D.W., \emph{Operator Algebras and Quantum Statistical
  Mechanics 2}, Texts and Monographs in Physics, Springer-Verlag, Berlin,
  second edn. (1997)

\bibitem{BFV}
Brunetti R., Fredenhagen K., Verch R., \enquote{{The Generally Covariant
  Locality Principle - A New Paradigm for Local Quantum Physics}}, Commun.
  Math. Phys. \textbf{237} (2003) 31, \eprint{math-ph/0112041}

\bibitem{Bu}
Buchholz D., \enquote{{On Hot Bangs and the Arrow of Time in Relativistic
  Quantum Field Theory}}, Commun. Math. Phys. \textbf{237} (2003) 271,
  \eprint{hep-th/0301115}

\bibitem{BuOjiRoo}
Buchholz D., Ojima I., Roos H., \enquote{{Thermodynamic Properties of
  Non-Equilibrium States in Quantum Field Theory}}, Annals Phys. \textbf{297}
  (2002) 219, \eprint{hep-ph/0105051}

\bibitem{BuSchl}
Buchholz D., Schlemmer J., \enquote{{Local Temperature in Curved Spacetime}},
  Class. Quant. Grav. \textbf{24} (2007) F25, \eprint{gr-qc/0608133}

\bibitem{EpGlaJaf}
Epstein H., Glaser V., Jaffe A., \enquote{Nonpositivity of the Energy Density
  in Quantized Field Theories}, Nuovo Cimento \textbf{36} (1965) 1016

\bibitem{GenWlKG}
Fewster C.J., \enquote{{A General Worldline Quantum Inequality}}, Class. Quant.
  Grav. \textbf{17} (2000) 1897, \eprint{gr-qc/9910060}

\bibitem{RevFewster}
Fewster C.J., \enquote{{Quantum Energy Inequalities and Stability Conditions in
  Quantum Field Theory}},   (2005), \eprint{math-ph/0502002}

\bibitem{FewOst}
Fewster C.J., Osterbrink L.W., \enquote{{Quantum Energy Inequalities for the
  Non-Minimally Coupled Scalar Field}}, J. Phys. \textbf{A41} (2008) 025402,
  \eprint{arXiv:0708.2450} \texttt{[gr-qc]}

\bibitem{GenWlEM}
Fewster C.J., Pfenning M.J., \enquote{{A Quantum Weak Energy Inequality for
  Spin-One Fields in Curved Spacetime}}, J. Math. Phys. \textbf{44} (2003)
  4480, \eprint{gr-qc/0303106}

\bibitem{FewRoNull}
Fewster C.J., Roman T.A., \enquote{{Null Energy Conditions in Quantum Field
  Theory}}, Phys. Rev. \textbf{D67} (2003) 044003, \eprint{gr-qc/0209036}

\bibitem{QEIvsWH2}
Fewster C.J., Roman T.A., \enquote{{On Wormholes with Arbitrarily Small
  Quantities of Exotic Matter}}, Phys. Rev. \textbf{D72} (2005) 044023,
  \eprint{gr-qc/0507013}

\bibitem{FewSmith}
Fewster C.J., Smith C.J., \enquote{{Absolute Quantum Energy Inequalities in
  Curved Spacetime}},   (2007), \eprint{gr-qc/0702056}

\bibitem{GenWlDirac}
Fewster C.J., Verch R., \enquote{{A Quantum Weak Energy Inequality for Dirac
  Fields in Curved Spacetime}}, Commun. Math. Phys. \textbf{225} (2002) 331,
  \eprint{math-ph/0105027}

\bibitem{CJF-RV}
Fewster C.J., Verch R., \enquote{{Stability of Quantum Systems at Three Scales:
  Passivity, Quantum Weak Energy Inequalities and the Microlocal Spectrum
  Condition}}, Commun. Math. Phys. \textbf{240} (2003) 329,
  \eprint{math-ph/0203010}

\bibitem{Ford}
Ford L.H., \enquote{Quantum Coherence Effects and the Second Law of
  Thermodynamics}, Proc. R. Soc. Lond. A \textbf{364} (1978) 227

\bibitem{QEIvsWH1}
Ford L.H., Roman T.A., \enquote{{Quantum Field Theory Constrains Traversable
  Wormhole Geometries}}, Phys. Rev. \textbf{D53} (1996) 5496,
  \eprint{gr-qc/9510071}

\bibitem{Fulling}
Fulling S.A., \enquote{{Nonuniqueness of Canonical Field Quantization in
  Riemannian Space-Time}}, Phys. Rev. \textbf{D7} (1973) 2850

\bibitem{GuidoLongo}
Guido D., Longo R., \enquote{{A Converse Hawking-Unruh Effect and dS(2)/CFT
  Correspondence}}, Ann. Henri Poinc. \textbf{4} (2003) 1169

\bibitem{HawkEllis}
Hawking S.W., Ellis G.F.R., \emph{The Large Scale Structure of Space-time},
  Cambridge University Press, London (1973)

\bibitem{HolWald-wick1}
Hollands S., Wald R.M., \enquote{{Local Wick Polynomials and Time Ordered
  Products of Quantum Fields in Curved Spacetime}}, Commun. Math. Phys.
  \textbf{223} (2001) 289, \eprint{gr-qc/0103074}

\bibitem{Hueb}
H{\"u}bener R., \emph{Lokale Gleichgewichtszust{\"a}nde massiver Bosonen},
  Diplomarbeit, University of G{\"o}ttingen (2005)

\bibitem{KayWald}
Kay B.S., Wald R.M., \enquote{Theorems on the Uniqueness and Thermal Properties
  of Stationary, Nonsingular, Quasifree States on Spacetimes with a Bifurcate
  Killing Horizon}, Phys. Rept. \textbf{207} (1991) 49

\bibitem{Moretti}
Moretti V., \enquote{{Comments on the Stress-Energy Tensor Operator in Curved
  Spacetime}}, Commun. Math. Phys. \textbf{232} (2003) 189,
  \eprint{gr-qc/0109048}

\bibitem{MorThorYurt}
Morris M.S., Thorne K.S., Yurtsever U., \enquote{{Wormholes, Time Machines, and
  the Weak Energy Condition}}, Phys. Rev. Lett. \textbf{61} (1988) 1446

\bibitem{QEIvsWarp}
Pfenning M.J., Ford L.H., \enquote{{The Unphysical Nature of `Warp Drive'}},
  Class. Quant. Grav. \textbf{14} (1997) 1743, \eprint{gr-qc/9702026}

\bibitem{GFRoman}
Roman T.A., \enquote{{On the 'Averaged Weak Energy Condition' and Penrose's
  Singularity Theorem}}, Phys. Rev. \textbf{D37} (1988) 546

\bibitem{RevRoman}
Roman T.A., \enquote{{Some Thoughts on Energy Conditions and Wormholes}}, in
  Novello M., Perez~Bergliaffa S., Ruffini R., eds., \enquote{Proceedings of
  the MG10 Meeting}, p. 1909, \eprint{gr-qc/0409090}

\bibitem{SahlVe}
Sahlmann H., Verch R., \enquote{{Passivity and microlocal spectrum condition}},
  Commun. Math. Phys. \textbf{214} (2000) 705, \eprint{math-ph/0002021}

\bibitem{CasimEng1}
Serry F., Walliser D., Maclay G.J., \enquote{The Role of the Casimir Effect in
  the Static Deflection and Stiction of Membrane Strips in
  Microelectromechanical Systems (MEMS)}, J. Appl. Phys. \textbf{84} (1998)
  2501

\bibitem{GFTipler}
Tipler F.J., \enquote{Energy Conditions and Spacetime Singularities}, Phys.
  Rev. \textbf{D17} (1978) 2521

\bibitem{Unruh}
Unruh W.G., \enquote{{Notes on Black Hole Evaporation}}, Phys. Rev.
  \textbf{D14} (1976) 870

\bibitem{VeHad}
Verch R., \enquote{{Local Definiteness, Primarity and Quasiequivalence of
  Quasifree Hadamard Quantum States in Curved Space-Time}}, Commun. Math. Phys.
  \textbf{160} (1994) 507

\bibitem{Wald78}
Wald R.M., \enquote{Trace Anomaly of a Conformally Invariant Quantum Field in
  Curved Space-Time}, Phys. Rev. \textbf{D17} (1978) 1477

\bibitem{WaldGR}
Wald R.M., \emph{General Relativity}, University of Chicago Press, Chicago, IL
  (1984)

\bibitem{WaldQFT}
Wald R.M., \emph{Quantum Field Theory in Curved Spacetime and Black Hole
  Thermodynamics}, Chicago Lectures in Physics, University of Chicago Press,
  Chicago, IL (1994)

\bibitem{YurWal}
Wald R.M., Yurtsever U., \enquote{{General Proof of the Averaged Null Energy
  Condition for a Massless Scalar Field in Two-dimensional Curved Space-
  Time}}, Phys. Rev. \textbf{D44} (1991) 403


\end{thebibliography}
\end{document}